\documentstyle[12pt]{article}

\newcommand{\bib}{\bibitem}
\newcommand{\be}{\begin{equation}}
\newcommand{\ee}{\end{equation}}
\newcommand{\er}{\end{eqnarray}}
\newcommand{\br}{\begin{eqnarray}}
\newcommand{\dslash}{\partial\!\!\!/}
\newcommand{\aslash}{A\!\!\!/}
\newcommand{\bslash}{B\!\!\!/}
\newcommand{\kslash}{\kappa\!\!\!/}
\newcommand{\fslash}{f\!\!\!/}
\newcommand{\Dslash}{D\!\!\!\!/}
\begin{document}

\thispagestyle{empty} 
$\phantom{x}$\vskip 0.618cm

\begin{center}
{\Huge {\bf Soldering Formalism: Theory and Applications\footnote{ Based on
work done in collaboration with R.Banerjee and E.M.C.Abreu}}} 
\vspace{1cm}
$\phantom{X}$\\
{\Large Clovis Wotzasek\footnote{Talk presented at
Universite de Montreal, Nov 24-27, 1997.}}\\
[3ex]{\em Instituto de F\'\i sica\\Universidade
Federal do Rio de Janeiro\\21945, Rio de Janeiro, Brazil\\}

\end{center}

\vspace{1cm}

\begin{abstract}
\noindent The soldering mechanism is a new technique to work with distinct
manifestations of dualities that incorporates interference
effects, leading to new physical results
that includes quantum contributions.
To work out this formalism in different scenarios we have
developed a systematic method of obtaining duality symmetric actions
in any space-time dimensions.
This approach was used to investigate the cases of
electromagnetic dualities, and $D\geq 2$ bosonization.
In the former context
this technique is applied for the quantum mechanical harmonic
oscillator, the scalar
field theory in two dimensions and the Maxwell theory in four dimensions.
In all cases there are two such distinct actions. Furthermore, by
soldering these
actions in any dimension a master action
is obtained which is duality invariant under a
much bigger set of symmetries than is usually envisaged. The concept of
swapping duality is introduced and its implications are discussed.
The effects of coupling to gravity are also elaborated. Finally, the
extension of the analysis for arbitrary dimensions is indicated.
In the later context, a technique is developed
that solders the dual aspects of some
symmetry following from the bosonisation of two distinct fermionic
models, thereby leading to new results which cannot be otherwise
obtained. Exploiting this technique, the two dimensional chiral
determinants with opposite chirality are soldered to reproduce
either the
usual gauge invariant expression leading to the Schwinger model or,
alternatively, the Thirring model. Likewise, two apparently
independent three dimensional massive Thirring models with same
coupling but opposite mass signatures, in the long wavelegth limit,
combine by the process of bosonisation and soldering
to yield an effective massive Maxwell theory.
The current bosonisation formulas are given, both in
the original independent formulation as well as the effective theory, and
shown to yield consistent results for the correlation functions.
Similar features
also hold for quantum electrodynamics in three dimensions.
\end{abstract}

\newpage

\section{Introduction}

The role of duality as a qualitative tool in the investigation
of physical systems is being gradually disclosed at different
context and dimensions\cite{AG}.
Much effort has been given
in sorting out several technical aspects of duality symmetric actions.
A new technique, able to work with
distinct manifestations of the duality symmetry was proposed
many years ago by Stone\cite{S}.
The Stones' soldering mechanism for fusing together opposite
aspects of duality symmetries, provides a new formalism that
includes the quantum interference effects between the independent
components.  This leads to a new and unique way of obtaining
physical results
that includes quantum contributions, and is dimension independent.

Electromagnetic duality
in 2D and 4D is reexamined under the soldering point of view providing
interesting new results, both explitly dual and
covariant as the result of the interference between
self and anti-self dual actions displaying opposite aspects of
the electromagnetic duality symmetry\cite{BW60}.
Even in the most elementary problem in theoretical physics,
the simple harmonic oscillator, the soldering technique has
been able to produce new and interesting results.  In fact it
is precisely this simple manisfestation of
duality that pervades all field theoretical
examples as will be explicitly shown.

Exciting new results in 2D and 3D bosonization were generated
via soldering that could not be obtained by any of the
known existent bosonization techniques.
In 2D this technique allowed us to show how massless chiral
fields combine to provide the gauge invariant massive mode
present in the $QED_2$, showing that the Schwinger
mechanism is the result of quantum interference between
right and left moving modes,
very much like in the optical double slit Young's experiment.
In fact, by equipping the soldering technique with
gauge and Bose symmetry\cite{RB3}, it automatically
selects, in a unique way, the massless sector of
the chiral models displaying
the Jackiw-Rajaraman parameter that reflects
the bosonization ambiguity by $a=1$.
In the 3D case, the soldering mechanism was used to show the
result of fusing together two topologically massive modes
generated by the bosonization of two
massive Thirring models with opposite mass signatures in the long
wave-length limit.  The bosonized modes, which are described by
self and anti-self dual Chern-Simons models\cite{TPN,DJ}, were
then soldered into
the two massive modes of the 3D Proca model\cite{BW}.

These Lecture Notes are divided into three sections.  we conclude
this first
section elaborating
over the soldering formalism and the ideas of self and anti-self
dualities.
This is illustrated further in section II with a very simple example,
the one-dimensional harmonic oscillator mentioned above.
Indeed this example
is worked out in some details to unveil the key concepts of our
approach and to set the general tone of the Notes.
In fact, the extension
to field theory is more a matter of technique rather
than introducing truly new and fundamental concepts.                          
The remaing of this section is devoted to the discussion of
electromagnetic dualities in 4D, but due to the special nature
of the four-dimensional duality transformation, it becomes
mandatory to discuss
the even dimensional case in its full generality.
Indeed, for this problem the dimensionality of space time appears to
be extremely crucial. To treat all even dimensional cases
in a unified way, one
introduces the idea of an internal two-dimensional space.             
This internal space will generalize the idea of dual operation
in such way that the concept of self and anti-self duality will
be well defined in all even $D=2k+2$ dimensions.
Their explicit realisation is one of
the central results of our work.

The analyse of special features and applications of bosonization
is the object of the third section of the Notes.
We show that two distinct models
displaying dual aspects of some symmetry can be combined by the
simultaneous implementation of bosonisation and soldering to yield a
completely new theory. Importantly, dimensionality poses no constraint
in this analyses.
We start studying the two
dimensional case where bosonisation is known to yield exact results. 
Using bosonised expressions, the soldering mechanism
fuses, in a precise way,  left and right chiralities. This leads to
a general lagrangean in which the chiral symmetry no longer exists, but
it contains  two extra parameters manifesting the bosonisation ambiguities.
It is shown that different  parametrisations lead to different
models. Our results are not restrict to
the gauge invariant Schwinger model but the Thirring
model is also reproduced. We take the opportunity to
stress the importance of Bose symmetry in the chiral bosonization
analysis.
Some interesting consequences regarding the arbitrary
parametrisation in the chiral Schwinger model are also charted.

In the subsequent section, the discussion of the
three dimensional bosonization illuminates the full power and utility of
the present approach. Recall that bosonisation in
higher dimensions is not
exact, nevertheless, for massive fermionic models in the large mass or,
equivalently, the long wavelength limit, well defined local expressions
are known to exist\cite{M,RB}. 
Interestingly, these expressions exhibit a self
or an anti self dual symmetry that is dictated by the signature of the fermion
mass. It is precisely these symmetries that simulates the dual aspects of
the left and right chiral symmetry in the two dimensions.
Indeed, two distinct
massive Thirring models with opposite mass signatures,
are soldered to yield a massive Maxwell theory.
A direct comparison of the current correlation functions
obtained before and after the soldering process are carefully
developed to confirm the utility and practical nature of this
new formalism.

\subsection{The Soldering Formalism}

The technique of soldering essentially comprises in lifting the gauging
of a global symmetry to its local version and exploits certain concepts
introduced in a different context by Stone\cite{S} and one of us\cite{W}.
The analysis is intrinsically
quantal without having any classical analogue. This is easily explained
by the observation that a simple addition of two independent classical
lagrangeans is a trivial operation without leading to anything meaningful
or significant.  On the other hand, the direct sum of classical actions
depending on different fields would not give anything new.  It is the
soldering process that leads to a new and nontrivial result.

The basic idea of the soldering procedure is to raise a global Noether
symmetry of the self and anti-seld dual constituents into a
local one, but for an effective composite system, consisting of the
dual components and an interference term.  This algorithm,
consequently, defines the
soldered action.  Here we shall adopt an iterative Noether procedure
to lift the global symmetries.  Therefore, assume that the symmetries
in question are being described by the local actions
$S_{\pm}(\phi_{\pm}^\eta)$,
invariant under a global multi-parametric transformation

\be
\label{ii10}
\delta \phi_{\pm}^\eta = \alpha^\eta
\ee
Here $\eta$ represents the
tensorial charachter of the basic fields in the dual actions
$S_{\pm}$ and, for notational simplicity, will be dropped from now on.
Now, under  local
transformations these actions will not remain invariant,
and Noether counter-terms
become necessary to restablish the invariance, along with
appropriate compensatory soldering fields $B^{(N)}$,
\be
\label{ii20}
S_{\pm}(\phi_{\pm})^{(0)}\rightarrow S_{\pm}(\phi_{\pm})^{(N)}=
S_{\pm}(\phi_{\pm})^{(N-1)}- B^{(N)} J_{\pm}^{(N)}
\ee
Here $J_{\pm}^{(N)}$ are the Noether currents, and (N) is the iteration
number. For the self and anti-self dual
systems we have in mind, this iterative gauging procedure is
(intentionally) construct
not to produce invariant actions for any finite number of steps.
However, if after N repetitions the non invariant piece end up
being only dependend on the gauging parameters, but not on the original
fields, there will exist the possibility of mutual cancelation, if both
self and anti-seld gauged systems are put together.   Then, suppose that
after N repetitions we arrive at the following
simultaneous conditions,

\br
\label{ii30}
\delta S_{\pm}(\phi_{\pm})^{(N)} \neq 0\nonumber\\
\delta S_{B}(\phi_{\pm})=0
\er
with
\be
\label{ii40}
S_{B}(\phi_{\pm})=S_{+}^{(N)}(\phi_{+}) + S_{-}^{(N)}(\phi_{-})+
\mbox{Contact Terms}
\ee
Then we can immediately identify the (soldering) interference term as,
\be
\label{ii50}
S_{int}=\mbox{Contact Terms}-\sum_{N}B^{(N)} J_{\pm}^{(N)}
\ee
where the Contact Terms are generally quadratic functions of the
soldering fields.  Incidentally, these auxiliary fields $B^{(N)}$
may be eliminated
from the resulting effective action, in favor of the physically relevant
degrees of freedom.
It is important to notice that after elimination of the soldering fields,
the resulting effective action
will not depend on either self or anti-self dual fields $\phi_{\pm}$
but only in some collective field, say $\Phi$,
defined in terms of the original ones in a (Noether) invariant way.

\be
\label{ii60}
S_B(\phi_{\pm})\rightarrow S_{eff}(\Phi)
\ee
Once such effective action has been established, the physical consequences
of the soldering are readily obtained by simple inspection.  This will
progressively be clarified in the especific applications to
be given in the
sections that follow.


\section{Duality Symmetry and Soldering in Different Dimensions}

In recent times the old idea \cite{O, GH, JS, Z, DT} of eletromagnetic
duality has been
revived with considerable attention and emphasis \cite{SS, GR, NB, DGHT}. 
Recent directions \cite{SS, KP, PST, G} also
include an 
abstraction of manifestly covariant forms  for such
actions or an explicit proof of their equivalence with the nonduality
symmetric actions, which they are supposed to represent. There are also
different suggestions on the possible analogies between duality symmetric
actions in different dimensions. In particular it has been claimed
\cite{PST} that
the two dimensional self dual 
action given in \cite{FJ} is the analogue of the four
dimensional electromagnetic duality symmetric action \cite{SS}.
In spite of the recent spate of papers on this subject there does not
seem to be a simple clear cut way of arriving at duality symmetric actions.
Consequently the fundamental nature of duality remains clouded by
technicalities. Additionally, the dimensionality of space time appears to
be extremely crucial. For instance, while the 
duality symmetry in $D=4k$ dimensions
is characterised by the one-parameter continuous group $SO(2)$, that in
$D=4k+2$ dimensions is described by a discrete group with just two
elements \cite{DGHT}.
Likewise, it has also been argued from general notions that a symmetry 
generator exists only in the former case. From an algebraic point of view
the distinction between the dimensionalities is manifested by the
following identities,
\br
\label{ssi1}
\mbox{}^{**}F &=& F\,\,\,;\,\,\,D=4k+2\nonumber\\
 &=&- F\,\,\,;\,\,\,D=4k
\er
where the $*$ denotes a usual Hodge 
dual operation and $F$ is the $\frac{D}{2}$-form.
Thus there is a self dual operation in the former which is missing in the
latter dimensions. This apparently leads to separate consequences for duality
in these cases.

The object of this section is to develop a method for systematically
obtaining and investigating different aspects of
duality symmetric actions that embrace all dimensions. A deep unifying
structure is illuminated which also leads to new symmetries. Indeed we
show that duality is not limited to field or string theories, but is
present even in the simplest of quantum mechanical examples- the harmonic
oscillator. It is precisely this duality which pervades all field
theoretical examples as will be explicitly shown. The basic idea of our
approach is deceptively simple. We start from the second order action for
any theory and
convert it to the first order form by introducing an auxiliary variable.
Next, a suitable relabelling of variables is done which induces an
internal index in the theory.  It is crucial to note that 
there are two distinct classes of
relabelling characterised by the opposite signatures of the
determinant of the $2\times 2$ orthogonal matrix defined in the internal 
space. 
Correspondingly, in this  space there are two  actions that are 
manifestly duality symmetric. 
Interestingly, their equations of motion are just the self and anti-self
dual solutions, where the dual field in the internal space 
is defined below in (\ref{ssi2}). It
is also found that in all cases there is one (conventional duality) 
symmetry transformation which preserves the invariance of these actions
but there is another transformation which swaps the actions. We refer
to this property as swapping duality. This
indicates the possibility, in any dimensions, 
of combining the two actions to a master action
that would contain all the duality symmetries.
Indeed this construction is explicitly done by
exploiting the ideas of soldering introduced in \cite{S} and developed by
us \cite{ABW, BW}. The soldered master action also  has manifest Lorentz
or general coordinate invariance. The generators of the symmetry
transformations are also obtained. 

It is easy to visualise how the internal space effectively 
unifies the results in the different $4k+2$
and $4k$ dimensions. The dual field is now defined to include the internal
index $(\alpha, \beta)$ in the fashion,
\br
\label{ssi2}
\tilde F^\alpha &=&\epsilon^{\alpha\beta}\mbox{}^{*}F^\beta 
\,\,\,;\,\,\,D=4k\nonumber\\
\tilde F^\alpha &=&\sigma_1^{\alpha\beta}\mbox{}^{*}F^\beta 
\,\,\,;\,\,\,D=4k+2
\er
where $\sigma_1$ is the usual Pauli matrix and 
$\epsilon_{\alpha\beta}$ is the fully
antisymmetric $2\times 2$ matrix with $\epsilon_{12} =1$. Now, 
irrespective of the
dimensionality, the repetition of the dual operation yields,
\be
\label{ssi3}
\tilde{\tilde F} = F
\ee
which generalises the relation (\ref{ssi1}). An immediate consequence of
this is the possibility to obtain self and anti-self dual solutions in
all even $D=2k+2$ dimensions. Their explicit realisation is one of
the central results of the paper.

This section is organised into five subsections. 
In section 2.1 the above ideas
are exposed by considering the example of the simple
harmonic oscillator. Provocativelly, a
close parallel with the electromagnetic notation is
also developed to illuminate the connection between this exercise and
those given for the field theoretical models in the next subsections. The
duality of scalar field theory in two dimensions is considered
in section 2.2.
The occurrence of a pair of actions is shown which exhibit duality and
swapping symmetries. These are the analogues of the four
dimensional electromagnetic duality symmetric actions.
Indeed, from these expressions, it is a trivial matter to
reproduce both the self and anti-self 
dual actions given in \cite{FJ}. Our analysis clarifies several issues
regarding the intertwining roles of chirality and duality in two dimensions.
The soldering of the pair of duality symmetric actions 
is also performed leading to fresh insights.
The analysis is completed by
including the effects of gravityin section 2.3. 
In section 2.4, the Maxwell theory is
treated in great details. Following our prescription the duality symmetric
action given in \cite{SS} is obtained. However, there is also a new
action which is  duality symmetric. Once again the soldering of these actions
leads to a master action which contains a much richer structure of
symmetries. Incidentally, it also manifests the original symmetry
that interchanges the Maxwell equations with the Bianchi identity, but
reverses the signature of the action.  In section 2.5,
the effects of gravity are
straightforwardly included. Section 2.6 contains the concluding comments.

\subsection {Duality in $0+1$ dimension}
The basic features of duality symmetric actions are already present
in the quantum mechanical examples as the present analysis on the
harmonic oscillator will clearly demonstrate. Indeed, this simple 
example is worked out in some details to illustrate the key concepts of
our approach and set the general tone of the paper. An extension to field
theory is more a matter of technique rather than introducing truly new
concepts. The
Lagrangean for the one-dimensional oscillator is given by,
\be
L=\frac{1}{2}\Big ({\dot q}^2-q^2\Big)
\label{ss10}
\ee
leading to an equation of motion,
\be
\ddot q+q=0
\label{ss20}
\ee
Introducing a change of variables,
\be
E=\dot q\,\,\,\,\,; \,\,\,\,\, B=q
\label{ss30}
\ee
so that,
\be
\dot B-E=0
\label{ss40}
\ee
is identically satisfied, the above equations (\ref{ss10}) and (\ref{ss20}) 
are, respectively, expressed as follows;
\be
L=\frac{1}{2}\Big(E^2-B^2\Big)
\label{ss50}
\ee
and,
\be
\dot E+B=0
\label{ss60}
\ee
It is simple to observe that  the transformations,\footnote{Note that
these are just the discrete cases ($\alpha=\pm\frac{\pi}{2}$) for a general
$SO(2)$ rotation matrix parametrised by the angle $\alpha$}
\be
E\rightarrow \pm B\,\,\,;\,\,\, B\rightarrow \mp E
\label{ss70}
\ee
swaps the equation of motion (\ref{ss60}) with the identity (\ref{ss40})
although the Lagrangean (\ref{ss50}) is not invariant. The similarity
with the corresponding analysis in the Maxwell theory is quite striking,
with $q$ and $\dot q$ simulating the roles of the magnetic and electric
fields, respectively. There is a duality among the equation of motion
and the `Bianchi' identity (\ref{ss40}), which is not
manifested in the Lagrangean.

In order to elevate the duality to the Lagrangean, the basic step is to
rewrite (\ref{ss10}) in the first order form by introducing an
additional variable,
\br
L&=&p\dot q-\frac{1}{2}(p^2+q^2)\nonumber\\
&=&\frac{1}{2}\Big(p\dot q-q\dot p -p^2-q^2\Big)
\label{ss80}
\er
where a symmetrisation has been performed. There are now
two possible classes for relabelling these variables corresponding to
proper and improper rotations generated by the matrices $R^+(\theta)$ and
$R^-(\varphi)$ with determinant $+1$ and $-1$, respectively,
\br
\left(\begin{array}{c}
q\\
p
\end{array}\right)
=
\left(\begin{array}{cc}
{\cos\theta} & {\sin\theta} \\
{-\sin\theta} &{\cos\theta} \end{array}\right)
\left(\begin{array}{c}
x_1\\
x_2
\end{array}\right)
\label{ssmatrix1}
\er
\br
\left(\begin{array}{c}
q\\
p
\end{array}\right)
=
\left(\begin{array}{cc}
{\sin\varphi} & {\cos\varphi} \\
{\cos\varphi} &{-\sin\varphi} \end{array}\right)
\left(\begin{array}{c}
x_1\\
x_2
\end{array}\right)
\label{ssmatrix}
\er
leading to the distinct Lagrangeans,
\br
L_\pm&=&\frac{1}{2}\Big(\pm x_\alpha\epsilon_{\alpha\beta}\dot x_\beta
-x_\alpha^2\Big)\nonumber\\
&=&{1\over 2}\left( \pm B_\alpha\epsilon_{\alpha\beta}E_\beta-B_\alpha^2
\right)
\label{ss100}
\er
where we have reverted back to the `electromagnetic' notation
introduced in (\ref{ss30}).  By these change
of variables an index $\alpha=(1, 2)$ has been introduced that
characterises a symmetry in this internal space, the complete details of
which will progressively become clear.
It is useful to remark that the above change of variables are succinctly
expressed as,
\br
q&=&x_1\,\,\,\,;\,\,\,\, p=x_2\nonumber\\
q&=&x_2\,\,\,\,;\,\,\,\, p=x_1
\label{ss90}
\er
by setting the angle $\theta=0$ or $\varphi =0$ in the rotation matrices 
(\ref{ssmatrix1}) and (\ref{ssmatrix}). Correspondingly,
the Lagrangean (\ref{ss80}) goes over to (\ref{ss100}). 
Now observe
that the above Lagrangeans (\ref{ss100}) 
are manifestly invariant under the continuous duality
transformations,
\be
\label{ss90a}
x_\alpha\rightarrow R^+_{\alpha\beta}x_\beta
\ee
which may be equivalently expressed as,
\br
E_\alpha&\rightarrow& R^+_{\alpha\beta}E_\beta\nonumber\\
B_\alpha&\rightarrow& R^+_{\alpha\beta}B_\beta
\label{ss100a}
\er
where $R^+_{\alpha\beta}$ is the usual $SO(2)$ rotation matrix 
(\ref{ssmatrix1}).
The generator of the infinitesimal symmetry transformation is given by,
\be
\label{ssgen}
Q^\pm =\pm\frac{1}{2}x_\alpha x_\alpha
\ee
so that the complete transformations (\ref{ss90a}) are generated as,
\br
x_\alpha \rightarrow x'_\alpha & =& e^{-i\theta Q} x_\alpha e^{i\theta Q}
\nonumber\\ 
&=& R_{\alpha\beta}^+(\theta)x_\beta
\label{ssfingen}
\er
This is easy to verify by using the basic symplectic brackets obtained
from (\ref{ss100}),
\be
\label{sssymplectic}
\{x_\alpha , x_\beta\}=\mp \epsilon_{\alpha\beta}
\ee
Parametrising the angle by $\theta=\frac{\pi}{2}$ the discrete
transformation is obtained,
\br
E_\alpha&\rightarrow& \epsilon_{\alpha\beta}E_\beta\nonumber\\
B_\alpha&\rightarrow& \epsilon_{\alpha\beta}B_\beta
\label{ss110}
\er
This is the parallel of the usual constructions 
done in the Maxwell theory to
induce a duality symmetry in the action. 

Let us now comment on an
interesting property, which is related to the existence of two distinct
Lagrangeans (\ref{ss100}), by replacing (\ref{ss100a}) with a new set of
transformations,
\br
E_\alpha&\rightarrow& R^-_{\alpha\beta}(\varphi)E_\beta\nonumber\\
B_\alpha&\rightarrow& R^-_{\alpha\beta}(\varphi)B_\beta
\label{ss110b}
\er
Notice that these transformations preserve the invariance of
the Hamiltonian following from either $L_+$ or $L_-$. 
Interestingly, the kinetic
terms in the Lagrangeans change signatures so that $L_+$ and $L_-$ 
swap into one another. This feature of duality swapping
will subsequently recur in a different context and has important
implications in higher dimensions.

The discretised version of (\ref{ss110b}) is obtained by setting 
$\varphi =0$,
\br
E_\alpha&\rightarrow& \sigma_1^{\alpha\beta}E_\beta\nonumber\\
B_\alpha&\rightarrow& \sigma_1^{\alpha\beta}B_\beta
\label{ss110a}
\er
It is precisely the $\sigma_1$ matrix that reflects the proper into
improper rotations,
\be
\label{sseverton}
R^+(\theta) \sigma_1=R^-(\theta)
\ee
which illuminates the reason behind the swapping of the Lagrangeans in
this example.

Since we have systematically developed a procedure for obtaining a duality
symmetric Lagrangean, it is really not necessary to show its equivalence
with the original Lagrangean, as was done in the Maxwell theory. 
Nevertheless, to complete the analogy, we show that (\ref{ss100}) 
reduces to (\ref{ss50}) or (\ref{ss10}) by using the equation of motion,
\be
x_\alpha=\pm\epsilon_{\alpha\beta}\dot x_\beta
\label{ss120}
\ee
which can be reexpressed as,
\be
B_\alpha=\pm\epsilon_{\alpha\beta} E_\beta
\label{ss130}
\ee
to eliminate one component (say the variables with label 2)
from (\ref{ss100}).
This immediately reproduces
(\ref{ss50}) while (\ref{ss110}) reduces to (\ref{ss70}).

An important point to stress is that there are actually two, and not one,
duality symmetric actions $L_\pm$ (\ref{ss100}), corresponding to the
signatures in the determinant of the transformation matrices.  
As shall be shown in 
subsequent sections this is also true for the scalar field theory in $1+1$
dimensions and the electromagnetic theory in $3+1$ dimensions. Usually, in
the literature, only one of these 
is highlighted while the other is not mentioned.
We now elaborate the implications of this property which will also be
crucial in discussing field theoretical models. In the coordinate
language these Lagrangeans correspond to two bi-dimensional chiral
oscillators rotating
in opposite directions. This is easily
verified either by looking at the classical equations of motion or by
examining the spectrum of the angular momentum operator,
\be
J_\pm=\pm\epsilon_{ij}x_i p_j = \pm H
\label{ss140}
\ee
where $H$ is the Hamiltonian of the usual harmonic oscillator.
In other words the two Lagrangeans manifest the dual aspects of rotational
symmetry in the two-dimensional internal space. 
Consequently it is possible to solder them by following the
general techniques elaborated in \cite{ABW, BW}. This soldering as well as
its implications are the subject of the remainder of this section.

The soldering mechansim, it must be recalled, is intrinsically an operation
that has no classical analogue. The crucial point is that the Lagrangeans
(\ref{ss100}) are now considered as functions of independent variables,
namely $L_+(x)$ and $L_-(y)$, instead of the same $x$. A naive
addition of the classical Lagrangeans with the same variable is of course
possible leading to a trivial result. If,
on the other hand, the Lagrangeans are functions of distinct variables,
a straightforward addition does not lead to any new information. The soldering
process precisely achieves this purpose. Consider the gauging of the
Lagrangeans under the following gauge transformations,
\be
\delta x_\alpha=\delta y_\alpha=\dot\eta_\alpha
\label{ss150}
\ee
Then the gauge variations are given by,
\be
\delta L_\pm(z)=\epsilon_{\alpha\beta}
\dot\eta_\alpha J_\beta^{(\pm)}(z)\,\,\,\,;\,\,
z=x,y
\label{ss160}
\ee
where the currents are defined by,
\be
\label{ss170}
J_\alpha^{(\pm)}(z)=\pm\dot z_\alpha+\epsilon_{\alpha\beta}z_\beta
\ee
Introducing a new field $B_\alpha$ transforming as,
\be
\delta B_\alpha =\epsilon_{\beta\alpha}\dot\eta_\beta
\label{ss180}
\ee
that will effect the soldering, it
is possible to construct a first iterated Lagrangean,
\be
\label{ss190}
L_\pm^{(1)}=L_\pm-B_\alpha J_\alpha^\pm
\ee
The gauge variation of (\ref{ss190}) is easily obtained,
\be
\label{ss200}
\delta L_\pm^{(1)}=-B_\alpha\delta J_\alpha^\pm
\ee
Using the above results we define a second iterated Lagrangean,
\be
\label{ss210}
L_\pm^{(2)}=L_\pm^{(1)}-\frac{1}{2}B^2_\alpha
\ee
which finally leads to a Lagrangean,
\be
\label{ss220}
L=L_+^{(2)}(x)+L_-^{(2)}(y)=L_+(x) + L_-(y) -B_\alpha\Big(J_\alpha^+ (x)+
J_\alpha^- (y)\Big)-B_\alpha^2
\ee
that is invariant under the complete set of transformations
(\ref{ss150}) and (\ref{ss180}),i.e.;
\be
\label{ss230}
\delta L=0
\ee
It is now possible to eliminate the auxiliary $B_\alpha$ field by using
the equation of motion, which yields,
\be
\label{ss240}
B_\alpha=-\frac{1}{2}\Big (J_\alpha^+(x)+J_\alpha^-(y)\Big)
\ee
Inserting this solution back into (\ref{ss220}), we obtain the final
soldered Lagrangean,
\be
\label{ss250}
L(w)=\frac{1}{4}\Big(\dot w_\alpha^2-w_\alpha^2\Big)
\ee
which is no longer a function of $x$ and $y$ independently, but only
on their
gauge invariant combination,
\be
\label{ss260}
w_\alpha=\left(x_\alpha-y_\alpha\right)
\ee
The soldered Lagrangean just corresponds to a simple bi-dimensional 
oscillator. Thus, by starting from two Lagrangeans which contained the
opposite aspects of a duality symmetry, it is feasible to combine them
into a single Lagrangean which has a richer symmetry. A similar
phenomenon also exists in the field theoretical examples, as shall be
shown subsequently. 

Let us now expose all the symmetries of the above
Lagrangean. It is most economically done by recasting this Lagrangean in
two equivalent forms,
\be
L=\Omega_\alpha^+\Omega_\alpha^- = \bar\Omega_\alpha^+\bar\Omega_\alpha^-
\label{ssomega}
\ee
where,
\br
\Omega_\alpha^\pm &=&\frac{1}{2} \Big(
\dot w_\alpha\pm\Lambda_{\alpha\beta}w_\beta\Big)\nonumber\\
\bar\Omega_\alpha^\pm &=&\frac{1}{2} \Big(\Lambda_{\alpha\beta}\dot w_\beta
\pm w_\alpha\Big)\nonumber\\
\Lambda_{\alpha\beta}&=&\Big(R_{\alpha\beta}^+\,\,,\,\,
R_{\alpha\beta}^- \Big)
\label{sslambda}
\er
Now the Lagrangean (\ref{ssomega}) is {\it manifestly} symmetric under the
following continuous dual transformations,
\be
w_\alpha\rightarrow R^\pm_{\alpha\beta}\,w_\beta
\label{ssdual}
\ee
The transformation involving $R^+$ is just the original symmetry
(\ref{ss100a}). Those involving the $R^-$ matrices are the new symmetries.
Recall that the latter transformations 
swapped the two independent Lagrangeans
$L_\pm$. The soldered Lagrangean contains both combinations and hence
manifests both these symmetries. The corresponding symmetry group is
therefore $O(2)$. This is a completely new phenomenon. It
also occurs in field theory with certain additional subtle features.

The generator of the infinitesimal transformations that leads to the
$SO(2)$ rotation in
(\ref{ssdual}) is given by,
\be 
\label{ssdualgen}
Q=w_\alpha \epsilon_{\alpha\beta} \pi_\beta
\ee
so that,
\br
\label{ssdualfingen}
w_\alpha\rightarrow w'_\alpha &=& e^{-i\theta Q}\, w_\alpha\, e^{i\theta
Q}\nonumber\\ 
&=& R^+_{\alpha\beta}(\theta)\, w_\beta
\er
which is verified by using the canonical brackets,
\be
\label{ssbrackets}
\{w_\alpha, \pi_\beta\}=\delta_{\alpha\beta}
\ee

It is worthwhile to point out the quantum nature of the above calculation
by rewriting (\ref{ss250}), after an appropriate scaling of variables,  
in the form of an identity,
\br
\label{ss270}
L(x-y)&=&L(x)+L(y)-2x_\alpha^+y_\alpha^-\nonumber\\
z_\alpha^\pm&=&\frac{1}{\sqrt 2}(\dot z_\alpha\pm \epsilon_{\alpha\beta}
z_\beta)
\er
This shows that the Lagrangean of the 
simple harmonic oscillator expressed in terms of
the ``gauge invariant" variables $w=x-y$ is not obtained by just adding the
independent contributions. Rather, there is a contact term which manifests
the quantum effect. Indeed, the above identity can be interpreted as the
analogue of the well known Polyakov-Weigman \cite{PW} identity in two
dimensional field theory. As our analysis shows, such identities will
always occur whenever dual aspects of some symmetry are being soldered
or fused to yield a composite picture, irrespective of the
dimensionality of space-time \cite{ABW}. 
In the Polyakov-Weigman case it was the
chiral symmetry whereas here it was the rotational symmetry. 

\subsection{The Scalar Theory in 1+1 Dimensions}

The ideas developed in the previous section are now implemented and
elaborated in $1+1$
dimensions. It is simple to realise that the scalar theory is a very
natural example. For instance, in these dimensions, there is no photon and
the Maxwell theory trivialises so that the electromagnetic field can be
identified with a scalar field. Thus all the results presented here can be
regarded as equally valid for the ``photon" field. 
Indeed the computations will also be presented in a very suggestive
notation which  illuminates the Maxwellian nature of the problem.
Consequently the
present analysis is an excellent footboard for diving into the actual
electromagnetic duality discussed in the next section. The effects of
gravity are easily included in our approach as shown in a separate subsection.

The Lagrangean for the free massless scalar field is given by,
\be
\label{ssw10}
{\cal L}=\frac{1}{2} \Big(\partial_\mu\phi\Big)^2
\ee
and the equation of motion reads,
\be
\label{ssw20}
\ddot\phi-\phi ''=0
\ee
where the dot and the prime denote derivatives with respect to time and
space components, respectively. Introduce, as before, a 
change of variables using electromagnetic symbols,
\be
E=\dot\phi\,\,\,\,;\,\,\,\, B=\phi '
\label{ssw30}
\ee
Obviously, $E$ and $B$ are not independent but constrained by the
identity, 
\be
\label{ssw40}
E'-\dot B=0
\ee
In these variables the equation of motion and the Lagrangean are expressed
as,
\br
\label{ssw50}
&\mbox{}&\dot E-B'=0\nonumber\\
&\mbox{}&{\cal L}=\frac{1}{2}\Big(E^2-B^2\Big)
\er
It is now easy to observe that the transformations,
\be
\label{ssw60}
E\rightarrow \pm B\,\,\,\,;\,\,\,\, B\rightarrow \pm E
\ee
display a duality between the equation of motion and the `Bianchi'-like
identity (\ref{ssw40}) but the Lagrangean changes its signature.
Note that there is a relative change in the signatures
of the duality transformations (\ref{ss70}) and (\ref{ssw60}), arising
basically from dimensional considerations. This symmetry coresponds to the
improper group of rotations.

To illuminate the close connection with the Maxwell formulation, we
introduce covariant and contravariant vectors with a Minkowskian metric
$g_{00}=-g_{11}=1$, 
\be
F_\mu=\partial_\mu\phi\,\,\,;\,\,\, F^\mu=\partial^\mu\phi
\label{ssw60a}
\ee
whose components are just the `electric' and
`magnetic' fields defined earlier,
\be
F_\mu=\Big(E, B\Big)\,\,\,\,;\,\,\,\, F^\mu=\Big(E,- B\Big)
\label{ssw60b}
\ee
Likewise, with the convention $\epsilon_{01}=1$, the dual field is defined,
\br
\mbox{}^* F_\mu&=&\epsilon_{\mu\nu}\partial^\nu\phi =\epsilon_{\mu\nu}F^\nu 
\nonumber\\
&=& \Big(-B, -E\Big)
\label{ssw60c}
\er
The equations of motion and the `Bianchi' identity are now expressed by
typical electrodynamical relations,
\br
\partial_\mu F^\mu&=&0\nonumber\\
\partial_\mu \mbox{}^* F^\mu&=&0
\label{ssw60d}
\er

To expose a
Lagrangean duality symmetry, the basic principle of our approach 
to convert the original second order form
(\ref{ssw50}) to its first order version and then invoke a relabelling of
variables to provide an internal index, is adopted. This is easily achieved
by first introducing an auxiliary field,
\be
\label{ssw70}
{\cal L}= PE-\frac{1}{2}P^2-\frac{1}{2}B^2
\ee
where $E$ and $B$ have already been defined. The following renaming of
variables corresponding to the proper and improper transformations (see
for instance (\ref{ssmatrix1}) and (\ref{ssmatrix})
or (\ref{ss90})) is used,
\br
\label{ssw80}
\phi&\rightarrow&\phi_1\nonumber\\
P&\rightarrow&\pm\phi_2'
\er
where we are just considering the discrete sets (\ref{ss90}) of the full
symmetry (\ref{ssmatrix1}) and (\ref{ssmatrix}). 
Then it is possible to recast (\ref{ssw70}) in the form,
\br
\label{ssw90} 
{\cal L}\rightarrow{\cal L}_\pm&=&\frac{1}{2}\Bigg[\pm
{\phi'}_\alpha\sigma_1^{\alpha\beta}\dot\phi_\beta -{\phi'}_\alpha^2\Bigg] 
\nonumber\\
&=&\frac{1}{2}\Bigg[\pm
B_\alpha\sigma_1^{\alpha\beta}E_\beta-B_\alpha^2\Bigg] 
\er
In the second line the
Lagrangean is expressed in terms of the electromagnetic variables.
This Lagrangean is
duality symmetric under the transformations of the basic scalar fields,
\be
\label{ssw100}
\phi_\alpha\rightarrow\sigma^{\alpha\beta}_1\phi_\beta
\ee
which, in the notation of $E$ and $B$, is given by,
\br
\label{ssw110}
B_\alpha &\rightarrow &\sigma^{\alpha\beta}_1B_\beta\nonumber\\
E_\alpha &\rightarrow &\sigma^{\alpha\beta}_1E_\beta
\er
It is quite interesting to observe that, contrary to the harmonic
oscillator example or the electromagnetic theory discussed in the 
next section, the transformation matrix in the $O(2)$ space is not
the epsilon, but rather a Pauli matrix. 
This result is in  agreement with that found from general algebraic
arguments \cite{SS, DGHT} which stated that for $d=4k+2$ dimensions
there is a discrete $\sigma_1$ symmetry. Observe that 
(\ref{ssw110}) is a manifestation of the
original duality (\ref{ssw60}) which was also effected by the same operation.
It is important to stress that the above symmetry is only
implementable at the discrete level. Moreover, since it is not connected
to the identity, there is no generator for this transformation.

To complete the picture, we also
mention that the following rotation,
\be
\phi_\alpha\rightarrow \epsilon_{\alpha\beta}\phi_\beta
\label{ssw120}
\ee
interchanges the Lagrangeans (\ref{ssw90}),
\be
{\cal L}_+\leftrightarrow {\cal L}_-
\label{ssw130}
\ee
Thus, except for a rearrangement of the the matrices generating the
various transformations, most features of
the simple harmonic oscillator example are perfectly retained. The crucial
point of departure is that now all these transformations are only discrete.
Interestingly, the master action constructed below lifts these symmetries
from the discrete to the continuous.

Let us therefore solder the two distinct Lagrangeans to
manifestly display the complete symmetries. Before doing this it is
instructive to unravel the self and anti-self dual aspects of these
Lagrangeans, which are essential to physically understand the soldering
process.  The equations of motion following from (\ref{ssw90}),  
in the language of the basic fields, is given by,
\be
\partial_\mu\phi_\alpha=\mp\sigma^1_{\alpha\beta}\epsilon_{\mu\nu}
\partial^\nu\phi_\beta
\label{ssdual1}
\ee
provided reasonable boundary conditions are assumed. Note that although
the duality symmetric Lagrangean is not manifestly Lorentz covariant, the
equations of motion possess this property. We will return to this aspect
again in the Maxwell theory.
In terms of a vector field $F_\mu^\alpha$ and its dual $\mbox{}^*
F_\mu^\alpha$ defined analogously to
(\ref{ssw60b}), (\ref{ssw60c}), the equation of motion is rewritten as,
\be
\label{ssmotion}
F_\mu^\alpha=\pm\sigma_1^{\alpha\beta}\mbox{}^* F_\mu^\beta=\pm\tilde
F_\mu^\alpha
\ee
where the generalised Hodge dual $(\tilde F)$ has been defined in
(\ref{ssi2}).
This explicitly reveals the self and anti-self dual nature of the
solutions in the combined internal and coordinate spaces. The result
can be extended to any $D=4k+2$ dimensions with suitable insertion of indices.

We now  solder the two Lagrangeans. 
This is best done by using the notation of the basic fields  
of the scalar theory. These Lagrangeans ${\cal L}_+$ and ${\cal L}_-$ 
are regarded as functions of the
independent scalar fields $\phi_\alpha$ and $\rho_\alpha$. 
Consider the gauging of the following symmetry,
\be
\delta\phi_\alpha =\delta\rho_\alpha=\eta_\alpha
\label{ssw140}
\ee
Following exactly the steps performed for the harmonic oscillator example
the final  Lagrangean analogous to (\ref{ss220}) is obtained,
\be
{\cal L}={\cal L}_+(\phi)+{\cal
L}_-(\rho)-B_\alpha\Big(J_\alpha^+(\phi) +J_\alpha^-(\rho)\Big)-B_\alpha^2
\label{ssw150}
\ee
where the currents are given by,
\be
J_\alpha^\pm(\theta)=\pm\sigma_{\alpha\beta}^1\dot\theta_\beta-{\theta'}_\alpha
\,\,\,;\,\,\,\theta=\phi\,\,,\,\,\rho
\label{ssw160}
\ee
The above Lagrangean is gauge invariant under the extended
transformations including (\ref{ssw140}) and,
\be
\delta B_\alpha =\eta_{\alpha}'
\label{ssw170}
\ee
Eliminating the auxiliary $B_\alpha$ field using the equations of motion,
the final soldered Lagrangean is obtained from (\ref{ssw150}),
\be
{\cal L}(\Phi)= \frac{1}{4}\partial_\mu\Phi_\alpha \partial^\mu\Phi_\alpha
\label{ssw180}
\ee
where, expectedly, this is now only a function of the gauge invariant
variable, 
\be
\Phi_\alpha=\phi_\alpha-\rho_\alpha
\label{ssw190}
\ee
This master Lagrangean possesses all the symmetries
that are expressed by the continuous
transformations, 
\be
\label{ss205}
\Phi_\alpha\rightarrow R^\pm_{\alpha\beta}(\theta)\Phi_\beta
\ee
The generator corresponding to the $SO(2)$ transformations is easily obtained,
\br
\label{ss206}
Q&=&\int dy \Phi_\alpha \epsilon_{\alpha\beta} \Pi_\beta\nonumber\\
\Phi_\alpha\rightarrow \Phi'_\alpha &=&e^{-i\theta Q} \Phi_\alpha e^{i\theta
Q}
\er
where $\Pi_\alpha$ is the momentum conjugate to $\Phi_\alpha$.
Observe that either the original symmetry in $\sigma_1$ or the swapping
transformations were only at the discrete
level. The process of soldering has lifted these
transformations from the discrete to the continuous form. It is equally
important to reemphasize that the master action now possesses the $SO(2)$
symmetry which is more commonly associated with four dimensional duality
symmetric actions, and not for two dimensional theories. Note that by
using the electromagnetic symbols, the Lagrangean can be 
displayed in a form which manifests the soldering effect of
the self and anti self dual symmetries (\ref{ssmotion}),
\be
\label{ss207}
{\cal L}=\frac{1}{8}\Big(F_\mu^\alpha+\tilde F_\mu^\alpha\Big)
\Big(F^\mu_\alpha-\tilde F^\mu_\alpha\Big)
\ee
where the generalised Hodge dual in $D=4k+2$ dimensions has been
defined in (\ref{ssi2}).

An interesting observation is now made. Recall that the original duality
transformation (\ref{ssw60}) switching equations of motion into Bianchi
identities may be rephrased in the internal space by,
\br
\label{ssbianchi}
E_\alpha &\rightarrow& \mp R^\pm_{\alpha\beta} B_\beta\nonumber\\
B_\alpha &\rightarrow& \mp R^\pm_{\alpha\beta} E_\beta
\er
which is further written directly in terms of the scalar fields,
\be
\label{ssbianchi1}
\partial_\mu \Phi_\alpha \rightarrow \pm R^\pm_{\alpha\beta}
\epsilon_{\mu \nu}\partial^\nu \Phi_\beta
\ee
It is simple to verify that under these transformations even 
the Hamiltonian for the theories
described by the Lagreangeans
${\cal L}_\pm$ (\ref{ssw90}) are not invariant. However the Hamiltonian
following from the master
Lagrangean (\ref{ssw180}) preserves this symmetry. The Lagrangean itself
changes 
its signature. This is the exact analogue of the original situation.
A similar phenomenon also
occurs in the electromagnetic theory. This completes the discussion on the
symmetries of the master Lagrangean.

It is now straightforward to give a Polyakov-Weigman type identity,
that relates the ``gauge invariant" Lagrangean with the non gauge invariant
structures, by reformulating (\ref{ssw180}) after a scaling of the fields
$(\phi, \rho)\rightarrow \sqrt 2(\phi, \rho)$,
\be
\label{ssw210}
{\cal L}(\Phi)= {\cal L}(\phi)+{\cal L}(\rho)-2\partial_+\phi_\alpha
\partial_-\rho_\alpha
\ee
where the light cone variables are given by,
\be
\label{sslc}
\partial_\pm =\frac{1}{\sqrt 2}(\partial_0 \pm \partial_1)
\ee

Observe that, as in the harmonic oscillator example, the gauge invariance
is with regard to the transformations introduced for the soldering of the
symmetries. Thus, even if the theory does not have a gauge symmetry in the
usual sense, the dual symmetries of the theory can simulate the effects of
the former. This leads to a Polyakov-Wiegman type identity which has an
identical structure to the conventional identity.

Before closing this sub-section, it may be useful to highlight some other
aspects of duality which are peculiar to two dimensions, as for instance,
the chiral symmetry. The interpretation of this symmetry with regard to
duality seems, at least to us,
to be a source of some confusion and controversy. As is well known a
scalar field in two dimensions can be decomposed into two chiral pieces,
described by Floreanini Jackiw (FJ) actions \cite{FJ}. 
These actions are sometimes
regarded \cite{PST} as the two dimensional analogues of the duality symmetric 
four dimensional
electromagnetic actions \cite{SS}. Such an interpretation is debatable
since the latter have the $SO(2)$ symmetry (characterised by an internal
index $\alpha$) which is obviously lacking in
the FJ actions. Our analysis, on the other hand, has shown how to
incorporate this symmetry in the two dimensional case. Hence we consider
the actions defined by (\ref{ssw90}) to be the true analogue of the duality
symmetric electromagnetic actions to be discussed later. 
Moreover, by solving the equations
of motion of the FJ action, it is not possible to recover the second order
free scalar Lagrangean, quite in contrast to the electromagnetic theory 
\cite{SS}.
Nevertheless, since the FJ actions are just the chiral components of the
usual scalar action, these must be soldered to reproduce this result.
But if soldering is possible, such actions must also display the self and
anti-self dual aspects of chiral symmetry. This phenomenon is now explored
along with the soldering process.

The two FJ actions defined in terms of the independent scalar fields
$\phi_+$ and $\phi_-$ are given by,
\be
\label{ssw220}
{\cal L}^{FJ}_\pm(\phi_\pm)=\pm\dot\phi_\pm\phi_\pm'-\phi_\pm'\phi_\pm'
\ee
whose equations of motion show the self and anti self dual aspects,
\be
\label{ssw230}
\partial_\mu\phi_\pm=\mp\epsilon_{\mu\nu}\partial^\nu\phi_\pm
\ee
A trivial application of the soldering mechanism leads to,
\br
{\cal L}(\Phi)&=&{\cal L}^{FJ}_+(\phi_+)+{\cal L}^{FJ}_-(\phi_-)+\frac{1}{8}
\Big(J_+(\phi_+)+ J_-(\phi_-)\Big)^2\nonumber\\
&=&\frac{1}{2}\partial_\mu\Phi\partial^\mu\Phi
\label{ssw240}
\er
where the currents $J_\pm$ and the composite field $\Phi$ are given by,
\br
J_\pm&=&2\Big(\pm\dot\phi_\pm-\phi_\pm'\Big)\nonumber\\
\Phi &=& \phi_+-\phi_-
\label{ssw250}
\er
Thus the usual scalar action is obtained in terms of the composite field. 
The previous analysis has, however, shown that each of the Lagrangeans
(\ref{ssw90}) are equivalent to the usual scalar theory. Hence these
Lagrangeans contain both chiralities desribed by the FJ actions
(\ref{ssw220}). However, in the internal space, ${\cal L}_\pm$ carry the
self and anti self dual solutions, respectively. This clearly illuminates
the ubiquitous role of chirality versus duality in the two dimensional
theories which has been missed in the literature simply because, following
conventional analysis in four dimensions \cite{DT, SS}, 
only one particular duality symmetric Lagrangean ${\cal L}_-$
was imagined to exist. 

\subsection{Coupling to gravity}
It is easy to extend the analysis to include gravity. This is most
economically done by using the language of electrodynamics already
introduced. The Lagrangean for the scalar field coupled to gravity is
given by,
\be
{\cal L}= \frac{1}{2}\sqrt{-g}g^{\mu\nu}F_\mu F_\nu
\label{ssw260}
\ee
where $F_\mu$ is defined in
(\ref{ssw60b}) and $g=\det g_{\mu\nu}$. Converting
the Lagrangean to its first order form, we obtain,
\be
\label{ssw270}
{\cal L}=P E
-\frac{1}{2\sqrt{-g}g^{00}}\Big(P^2+B^2\Big)+\frac{g^{01}}{g^{00}} P B
\ee
where the $E$ and $B$ fields are defined in (\ref{ssw30}) and $P$ is an
auxiliary field. Let us next invoke a change of variables mapping
$(E, B)\rightarrow (E_1, B_1)$ by means of the
$O(2)$ transformation analogous to (\ref{ssw80}),
and relabel the variable $P$ by $\pm B_2$.
Then the Lagrangean (\ref{ssw270}) assumes the distinct forms,
\be
\label{ssw280}
{\cal L}_\pm=
\frac{1}{2}\Bigg[\pm B_\alpha\sigma_{\alpha\beta}^1 E_\beta-\frac{1}
{\sqrt{-g}g^{00}} B_\alpha^2\pm \frac{g^{01}}{g^{00}}
\sigma_{\alpha\beta}^1 B_\alpha B_\beta\Bigg]
\ee
which are duality symmetric under the transformations (\ref{ssw110}). 
As in the flat metric, there is a swapping between ${\cal L}_+$
and ${\cal L}_-$ if the transformation matrix is $\epsilon_{\alpha\beta}$.
To obtain a duality symmetric action for all 
transformations it is necessary to construct the master action
obtained by soldering the two independent pieces. The dual aspects of the
symmetry that will be soldered are revealed by looking at the equations of
motion following from (\ref{ssw280}),
\be
\label{ssw290}
\sqrt{-g}F_\mu^\alpha=\mp g_{\mu\nu}\sigma_1^{\alpha\beta}\mbox{}^*
F^{\nu, \beta} \ee
The result of the soldering process, 
following from our standard techniques, leads to the master Lagrangean,
\be
\label{ssw310}
{\cal L}=\frac{1}{4}\sqrt{-g}g^{\mu\nu}F_\mu^\alpha F_\nu^\alpha
\ee
where $F_\mu^\alpha$ is defined in terms of the composite field given in
(\ref{ssw190}).
In the flat space this just reduces to the expression found previously in
(\ref{ssw180}). It may be pointed out that, originating from this master
action it is possible, by passing to a first order form, to recover the
original pieces. 

To conclude, we show how the FJ action now follows trivially by taking any
one particular form of the two Lagrangeans, say ${\cal L}_+$. To make
contact with the conventional expressions quoted in the literature
\cite{SO}, it is
useful to revert to the scalar field notation, so that,
\be
{\cal L}_+= \frac{1}{2}\Bigg[\phi_1'\dot\phi_2+\phi_2'\dot\phi_1
+2\frac{g^{01}}{g^{00}}\phi_1'\phi_2'-\frac{1}{g^{00}\sqrt{-g}}\phi_\alpha'
\phi_\alpha'\Bigg]
\label{ssw320}
\ee
This is diagonalised by the following choice of variables,
\br
\phi_1&=&\phi_+ +\phi_-\nonumber\\
\phi_2 &=&\phi_+ - \phi_-
\label{ssw330}
\er
leading to,
\be
\label{ssw340}
{\cal L}_+={\cal L}_+^{(+)} (\phi_+, {\cal G}_+)
+{\cal L}_+^{(-)} (\phi_-, {\cal G}_-)
\ee
with,
\br
\label{ssw350}
{\cal L}_+^{(\pm)} (\phi_\pm, {\cal G}_\pm)&=&\pm\dot\phi_\pm\phi_\pm'
+{\cal G}_\pm\phi_\pm'\phi_\pm'\nonumber\\
{\cal G}_\pm &=&\frac{1}{g^{00}}\Bigg(-\frac{1}{\sqrt{-g}}\pm g^{01}\Bigg)
\er
These are the usual FJ actions in curved space as given in \cite{SO}. Such
a structure was suggested by gauging the conformal symmetry of the free
scalar field and then confirmed by checking the classical invariance under
gauge and affine transformations \cite{SO}. Here we have derived this
result directly from the action of the scalar field minimally coupled to
gravity. 

Observe that the explicit diagonalisation carried out in (\ref{ssw320}) for
two dimensions is actually a
specific feature of $4k+2$ dimensions. This is related to the basic identity
(\ref{ssi1}) governing the dual operation. 
If, however, one works with the master (soldered)
Lagrangean, then diagonalisation is possible in either $D=4k+2$ or
$D=4k$ dimensions since the corresponding identity (\ref{ssi3})
always has the correct signature.

\subsection{The Electromagnetic Duality}

Exploiting the ideas elaborated in the previous sections, it is straightforward
to implement duality in the electromagnetic theory. Let us start with
the usual Maxwell Lagrangean,
\be
{\cal L}=-\frac{1}{4}F_{\mu\nu}F^{\mu\nu}
\label{ssm10}
\ee
which is expressed in terms of the electric and magnetic fields
as,\footnote{Bold face letters denote three vectors.}
\be
{\cal L}= \frac{1}{2}\Big(\bf E^2-\bf B^2\Big)
\label{ssm20}
\ee
where,
\br
E_i&=&-F_{0i}=-\partial_0 A_i+\partial_i A_0\nonumber\\
B_i&=&\epsilon_{ijk}\partial_j A_k
\label{ssm30}
\er
The following duality transformation,
\be
\bf E\rightarrow \mp\bf B\,\,\,\,;\,\,\,\,\bf B\rightarrow \pm \bf E
\label{ssm40}
\ee
is known to preserve the invariance of the full set comprising
Maxwell's equations and the Bianchi identities although the Lagrangean
changes its signature. To have a duality symmetric Lagrangean, we now
know how to proceed in a systematic manner. The Maxwell Lagrangean
is therefore recast in a symmetrised first order form,
\be
{\cal L}=\frac{1}{2}\Big(\bf P.\dot{\bf A}-\dot{\bf P}.\bf A\Big)
-\frac{1}{2}{\bf P}^2-\frac{1}{2}\bf B^2
+A_0\bf\nabla.\bf P
\label{ssm50}
\ee
Exactly as was done for the harmonic oscillator, a change of
variables is invoked. Once again there are two possibilities
which translate
from the old set $(\bf P, \bf A)$ to the new ones $(\bf A_1, \bf A_2)$.
It is, however, important to recall that the Maxwell theory has a constraint
that is implemented by the Lagrange multiplier $A_0$. The redefined variables
are chosen which solve this constraint so that,
\br
\bf P&\rightarrow& \bf B_2\,\,\,\,;\,\,\,\,\bf A\rightarrow \bf A_1\nonumber\\
\bf P&\rightarrow& \bf B_1\,\,\,\,;\,\,\,\,\bf A\rightarrow \bf A_2
\label{ssm60}
\er
It is now simple to show that, in terms of the redefined variables, the
original Maxwell Lagrangean takes the form,
\be
{\cal L}_\pm={1\over 2}\left(\pm\bf {\dot A}_\alpha
\epsilon_{\alpha\beta}\bf B_\beta
-\bf B_\alpha\bf B_\alpha\right)
\label{ssm70}
\ee
Adding a total derivative that would leave the equations of motion unchanged,
this Lagrangean is expressed directly in terms of the electric and magnetic
fields,
\be
{\cal L}_\pm={1\over 2}\left(\pm\bf B_\alpha
\epsilon_{\alpha\beta}\bf E_\beta
-\bf B_\alpha\bf B_\alpha\right)
\label{ssm70a}
\ee
It is duality symmetric under the full $SO(2)$
transformations mentioned in an earlier context. Note that one of the
above structures (namely, ${\cal L}_-$) was given earlier in \cite{SS}.
Once again, in analogy with the harmonic oscillator example, it is
observed that the transformation (\ref{ss110b}) involving the $R^-$ matrices 
switches the Lagrangeans ${\cal L}_+$ and ${\cal L}_-$ into one another.
The generators of the $SO(2)$ rotations are given by,
\be
\label{ssjohn}
Q^{(\pm)}=\mp\frac{1}{2}\int d^3x\,\, {\bf{A}}^\alpha\,.\,{\bf{B}}^\alpha
\ee
so that,
\be
\label{ssgt}
{\bf{A}}_\alpha\rightarrow {\bf{A}}'_\alpha=e^{-iQ\theta}{\bf {A}}_\alpha
e^{iQ\theta} 
\ee
This can be easily verified by using the basic brackets following
from the symplectic structure of the theory,
\be
\label{ssgirotti}
\Big [A^i_\alpha(x),
\epsilon^{jkl}\partial^k A^l_\beta(y)\Big]=\pm i\delta^{ij}
\epsilon_{\alpha\beta} \delta({\bf{x}}-{\bf{y}})
\ee
It is useful to digress on the significance of the above analysis. Since
the duality symmetric Lagrangeans have been obtained directly from
the Maxwell Lagrangean, it is redundant to show the equivalence of
the former expressions with the latter, which is an essential
perquisite in other approaches. Furthermore, since classical
equations of motion have not been used at any stage, the purported
equivalence holds at the quantum level. The need for any explicit
demonstration of this fact, which has been the motivation of several
recent papers, becomes, in this analysis, superfluous. 
A related observation is that the usual way of showing the classical
equivalence is to use the
equations of motion to eliminate one component from (\ref{ssm70}),
thereby leading to the Maxwell Lagrangean in the temporal $A_0=0$ gauge.
This is not surprising since the change of variables leading from the
second to the first order form solved the Gauss law thereby
eliminating the multiplier. Finally, note that there are 
two distinct structures for the duality symmetric
Lagrangeans. These must correspond to the opposite aspects of some
symmetry, which is next unravelled. By looking at the equations of
motion obtained from (\ref{ssm70}),
\be
\bf {\dot A}_\alpha =
 \pm\epsilon_{\alpha\beta}\bf \nabla \times \bf A_\beta
\label{ssm80}
\ee 
it is possible to
verify that these are just the 
self and anti-self dual solutions,
\be
F_{\mu\nu}^\alpha=\pm\epsilon^{\alpha\beta}\mbox{}^* F_{\mu\nu}^\beta
\,\,;\,\,\mbox{}^* F_{\mu\nu}^\beta=\frac{1}{2}\epsilon_{\mu\nu\rho\lambda}F^{\rho
\lambda}_\beta
\label{ssm90}
\ee
obtained by setting $A_0^\alpha=0$.  Recall that in the two dimensional
theory the equation of motion naturally assumed a covariant structure.
Here, on the other hand, the introduction of $A_0^\alpha$ is necessary
since this term gives a vanishing contribution to the Lagrangean. This
feature distinguishes a gauge theory from the non gauge theory discussed
in the two dimensional example.
It may be observed that the opposite aspects
of the dual symmetry are contained in the internal space. Following our
quantum mechanical analogy, the next task is to solder the two Lagrangeans
(\ref{ssm70}). Consider then the gauging of the following symmetry,
\be
\delta \bf H_\alpha={\bf{h}}_\alpha \,\,\,;\,\,\,\bf H=\bf P, \bf Q
\label{ssm100}
\ee
where $\bf P$ and $\bf Q$ denote the basic fields in the Lagrangeans
${\cal L}_+$ and ${\cal L}_-$, respectively. The Lagrangeans
transform as,
\be
\delta {\cal L}_\pm=\epsilon_{\alpha\beta} 
\Big(\bf \nabla\times{\bf{h}}_\alpha\Big).\bf J_\beta^\pm
\label{ssm110}
\ee
with the currents defined by,
\be
\bf J_\alpha^\pm(\bf H)=\Big(\mp\bf \dot{H}_\alpha+\epsilon_{\alpha\beta}
\bf \nabla\times\bf H_\beta\Big)
\label{ssm120}
\ee
Next, the soldering  field $\bf W_\alpha$ is introduced which transforms as,
\be
\delta\bf W_\alpha =-\epsilon_{\alpha\beta}\bf\nabla\times{\bf{h}}_\beta
\label{ssm130}
\ee
Following standard steps as outlined previously, the final  Lagrangean
which is invariant under the complete set of transformations (\ref{ssm100})
and (\ref{ssm130}) is obtained,
\be
{\cal L}={\cal L}_+({\bf P})+{\cal L}_-({\bf Q})-{\bf W}^\alpha\,.\,\Big(\bf
J_\alpha^+ (\bf P)+ J_\alpha^- (\bf Q)\Big)- {\bf W}_\alpha^2
\label{ssm130a}
\ee
Eliminating the soldering field by using the
equations of motion, the effective
soldered Lagrangean following from (\ref{ssm130a}) is derived,
\be
{\cal L}=\frac{1}{4}\Bigg(\bf {\dot G}_\alpha.{\dot G}_\alpha
-\bf\nabla\times\bf G_\alpha.\bf\nabla\times\bf G_\alpha\Bigg)
\label{ssm140}
\ee
where the composite field is given by the  combination,
\be
\bf G_\alpha=\bf P_\alpha-\bf Q_\alpha
\label{ssm150}
\ee
which is invariant under (\ref{ssm100}).
It is interesting to note that, reinstating the $G_0^\alpha$ variable, 
this is nothing but the Maxwell Lagrangean
with a doublet of fields,
\be
{\cal L}=-\frac{1}{4} G_{\mu\nu}^\alpha G^{\mu\nu}_\alpha\,\,\,;\,\,\,
G_{\mu\nu}^\alpha=\partial_\mu  G_\nu^\alpha-
\partial_\nu  G_\mu^\alpha
\label{ssm160}
\ee
In terms of the original $ P$ and $Q$ fields it is once again possible,
like the harmonic oscillator example, to write  a Polyakov-Weigman like
identity,
\br
{\cal L}(P-Q)&=&{\cal L}(P)+{\cal L}(Q)-2 W_{i, \alpha}^+( P)
W_{i, \alpha}^-( Q)\nonumber\\
W_{i, \alpha}^\pm(H) &=&\frac{1}{\sqrt 2}\Big(F_{0i}^\alpha(H)
\pm \epsilon_{ijk}\epsilon_{\alpha\beta}
F_{jk}^\beta(H)\Big)\,\,\,;\,\,\, H=P, Q
\label {m170}
\er
With respect to the gauge transformatins (\ref{ssm100}), the above identity
shows that a contact term is necessary to restore the gauge invariant action
from two gauge variant forms. This, it may be recalled, is just the basic 
content of the Polyakov-Weigman identity. It is interesting to note that
the ``mass" term appearing in the above identity is composed of parity
preserving pieces $W_{i, \alpha}^\pm$, thanks to the presence of the
compensating $\epsilon$-factor from the internal space.

Following the oscillator example, it is now possible to show that by reducing
(\ref{ssm160}) to a first order from, we exactly obtain the two types of
the duality symmetric Lagrangeans
(\ref{ssm70a}). This shows the equivalence of
the soldering and reduction  processes.

A particularly illuminating way of rewriting the
Lagrangean (\ref{ssm160}) is,
\br
{\cal L} &=&-\frac{1}{8}\Big( G_{\mu\nu}^\alpha + \epsilon^{\alpha\beta}
\mbox{}^* G_{\mu\nu}^\beta\Big)
\Big( G^{\mu\nu}_\alpha -\epsilon_{\alpha\rho}
\mbox{}^* G^{\mu\nu}_\rho\Big)\nonumber\\
&=&-\frac{1}{8}\Big( G_{\mu\nu}^\alpha + 
\tilde G_{\mu\nu}^\alpha\Big)
\Big( G^{\mu\nu}_\alpha -\tilde G^{\mu\nu}_\alpha\Big)
\label{ssm180}
\er
where, in the second line, the generalised Hodge dual in the space 
containing the internal index has been used
to explicitly show the soldering of the self and anti self dual
solutions. A similar situation prevailed in the two dimensional analysis.
The above Lagrangean manifestly displays the following duality symmetries,
\be
A_{\mu}^\alpha \rightarrow R_{\alpha\beta}^\pm A_{\mu}^{\beta}
\label{ssm190}
\ee

\noindent where, without any loss of generality, 
we may denote the composite field,
of which $G_{\mu\nu}$ is a function, by $A$. 
The generator of the $SO(2)$ rotations is now given by,
\be
\label{ssm191}
Q=\int d{\bf{x}}\,\,
\epsilon^{\alpha\beta}{\bf {\Pi}}^\alpha\,\,.\,\,{\bf {A}}^\beta
\ee

Now observe that the master Lagrangean was
obtained from the soldering of two distinct Lagrangeans (\ref{ssm70}). The
latter were duality symmetric under both $\bf
A_\alpha\rightarrow\pm\epsilon_{\alpha\beta} \bf A_\beta$, while the
transformations involving the $\sigma_1$ matrix interchanged ${\cal L}_+$
with ${\cal L}_-$. The soldered
Lagrangean is therefore duality symmetric under the transformations 
(\ref{ssm190}). Furthermore, the discrete transformation related to the
$\sigma_1$ matrix has been lifted to its continuous form $R^-$.
The master Lagrangean, therefore, contains a bigger set of duality
symmetries than (\ref{ssm70}) and, significantly, is also manifestly
Lorentz invariant. Furthermore, recall 
that under the transformations mapping the field
to its dual, the original 
Maxwell equations are invariant but the  Lagrangean changes its signature.
The corresponding transformation in the $SO(2)$ space is
given by, 
\be
\label{ss0}
G_{\mu\nu}^\alpha \rightarrow R^+_{\alpha\beta}
\mbox{}^*G_{\mu\nu}^\beta
\ee
which, written in component notation, looks like,
\be
\bf E^\alpha\rightarrow\mp\epsilon^{\alpha\beta}\bf B^\beta\,\,\,;\,\,\,
\bf B^\alpha\rightarrow\pm\epsilon^{\alpha\beta}\bf E^\beta
\label{ssm200}
\ee
The standard duality symmetric Lagrangean fails to manifest this property.
However, as may be easily checked, the equations of motion obtained from 
the master Lagrangean swap with the corresponding Bianchi identity
while the Lagrangean flips sign. In this manner the original property
of the second order Maxwell Lagrangean is retrieved.
Note furthermore that the master Lagrangean possesses the $\sigma_1$ symmetry
(which is just the discretised version of $ R^-$), a feature expected for two
dimensional theories. A similar phenomenon occurred in the previous
section where the master action in two dimensions revealed the $SO(2)$
symmetry usually associated with four dimensional theories.

\subsection{Coupling to gravity}

To discuss how the effects of  gravity are included, we will proceed as in
the two dimensional example. The starting point is the Maxwell Lagrangean
coupled to gravity,
\be
{\cal
L}=-\frac{1}{4}\sqrt{-g}g^{\mu\alpha}g^{\nu\beta}F_{\mu\nu}F_{\alpha\beta}
\label{ssg10} 
\ee
>From our experience in the usual Maxwell theory we know that an eventual 
change of variables eliminates the Gauss law so that the term involving the
multiplier $A_0$ may be ignored from the outset. Expressing (\ref{ssg10}) in
terms of its components to separate explicitly the first and second order
terms, we find,
\be
\label{ssg20}
{\cal L}=\frac{1}{2}\dot A_i M^{ij} \dot A_j + M^i \dot A_i +M
\ee
where,
\br
M^{ij}&=&\sqrt{-g}\Big(g^{0i}g^{0j}- g^{ij}g^{00}\Big)\nonumber\\
M^{i}&=&\sqrt{-g} g^{0k}g^{ji} F_{jk}
\nonumber\\
M&=&\frac{1}{4}\sqrt{-g} g^{ij}g^{km} F_{im}F_{kj}
\label{ssg30}
\er
Now reducing the Lagrangean to its first order form, we obtain,
\be
\label{ssg40}
{\cal L}= P^i E_i-\frac{1}{2}P^i M_{ij} P^j -\frac{1}{2}M^i M_{ij} M^j
+P^i M_{ij} M^j +M
\ee
where $\dot A_i$ has been replaced by $E_i$ and 
$M_{ij}$ is the inverse of $M^{ij}$,
\be
\label{ssg50}
M_{ij}= \frac{-1}{\sqrt{-g}g^{00}} g_{ij}
\ee
with,
\be
\label{ssg60}
g^{\mu\nu}g_{\nu\lambda}=\delta^\mu_\lambda
\ee
Next, introducing the standard change of variables which solves the Gauss
constraint,
\br
\label{ssg70}
E_i &\rightarrow & E_i^{(1)}\nonumber\\
P^i&\rightarrow &\pm B^{i(2)}
\er
the Lagrangean (\ref{ssg40}) is expressed in the desired form,
\br
\label{ssg80}
{\cal L}_\pm=&\pm & E_i^\alpha\epsilon^{\alpha\beta}B_\beta^i
+\frac{1}{\sqrt{-g}g^{00}}g_{ij} B^i_\alpha B_\beta^j\nonumber\\
&\pm & \frac {g^{0k}}{g^{00}} \epsilon_{ijk}
\epsilon^{\alpha\beta}B_\alpha^i B_\beta^j
\er

Once again there are two duality symmetric actions corresponding to ${\cal
L}_\pm$. The enriched nature of the duality and swapping 
symmetries under a bigger set
of transformations, the constructing of a master Lagrangean from
soldering of ${\cal L}_+$ and ${\cal L}_-$, the corresponding
interpretations, all go through exactly as in the flat metric case.
Incidentally, the structure for ${\cal L}_-$ only was previously given in
\cite {SS}. 

\subsection{Final Discussions}
The ideas developed in these sections
revealed a unifying structure behind the construction of
the various duality symmetric actions. The essential ingredient was the
conversion of the second order action into a first order form followed by
an appropriate redefinition of variables such that these may be denoted
by  an internal index. The duality naturally occurred in this internal
space. Since the duality symmetric actions were directly derived from the
original action the proof of their equivalence becomes superfluous. This
is otherwise essential where such a derivation is lacking and recourse is
taken to either equations of motion  or some hamiltonian analysis.
Obviously the most simple and fundamental manifestation of the duality
property was in the context of the quantum mechanical harmonic oscillator.
Since a field is interpreted as a collection of an infinite set of such
oscillators, it is indeed expected and not at all surprising that all
these concepts and constructions are {\it almost} carried over entirely  
for field theories. It may be remarked that the extension of the harmonic
oscillator analysis to field theories has proved useful in other contexts
and in this particular case has been really clinching. Furthermore, by
invoking a highly suggestive electromagnetic notation for the harmonic
oscillator analysis, its close correspondence with the field theory
examples was highlighted.

A notable feature of the analyis was the revelation of a whole class of
new symmetries and their interrelations. Different aspects of this feature were
elaborated.  To be precise, it was shown that there are actually two
\footnote {Note that usual discussions of duality symmetry consider only
one of these actions, namely ${\cal L}_-$.}
duality symmetric actions $({\cal L}_\pm)$ 
for the same theory. These actions carry the opposite (self and anti self
dual) aspects of some symmetry and their occurrence  was
essentially tied to the fact that there were two distinct classes in which
the renaming of variables was possible, depending on the signature of the
determinant specifying the proper or improper rotations. 
To discuss further the implications of this pair of duality symmetric
actions it is best to compare with the existing results. This also serves
to put the present work in a proper perspective. It should be mentioned
that the analysis for two and four dimensions are generic for $4k+2$ 
and $4k$ dimensions, respectively.

It is usually observed \cite{DGHT} 
that the invariance of the actions in different
$D$-dimensions is preserved by the following groups,
\be
\label{ssc1}
{\cal G}_d=Z_2 \,\,\,;\,\,\, D=4k+2
\ee
and,
\be
\label{ssc2}
{\cal G}_c= SO(2)\,\,\,;\,\,\, D=4k
\ee
which are called the ``duality groups". The $Z_2$ group is a discrete
group with two elements, the trivial identity and the $\sigma_1$ matrix.
Observe an important difference
since in one case this group is continous while in the other it is
discrete. In our exercise this was easily verified by the pair of duality
symmetric actions ${\cal L}_\pm$. The new ingredient is that nontrivial
elements of these
groups are also responsible for the swapping ${\cal L}_+\leftrightarrow {\cal
L}_-$, but in the other dimensions. Thus the ``duality swapping matrices"
$\Sigma_s$ are given by,
\br
\label{ssc3}
{\Sigma}_s &=&\sigma_1\,\,\,;\,\,\, D=4k\nonumber\\
 &=& \epsilon\,\,\,;\,\,\, D=4k+2
\er

It was next shown that ${\cal L}_\pm$ contained the self and anti-self
dual aspects of some symmetry. Consequently, following the ideas developed
in \cite{ABW, BW}, the two Lagrangeans could be soldered to yield a master
Lagrangean ${\cal L}_m={\cal L}_+\oplus {\cal L}_-$. The master action, in
any dimensions, was manifestly Lorentz or general coordinate invariant
and was also duality symmetric under both the groups mentioned
above. Moreover the process of soldering lifted the discrete group $Z_2
$ to its continuous version. The duality group for the master action
in either dimensionality therefore simplified to,
\be
\label{sscm}
{\cal G}= O(2)\,\,\,;\,\,\, D=2k+2
\ee
Thus, at the level of the master action, the fundamental distinction 
between the odd and even $N$-forms gets obiliterated. It ought to be
stated that the  lack of usual Chern Simons terms in $D=4k+2$ dimensions
to act as the generators of duality transformations is compensated by the
presence of a similar term in the internal space. Thanks to this it was
possible to  explicitly construct the symmetry generators for the master
action in either two or four dimensions.

We also showed that the master actions in any dimensions, apart from
being duality symmetric under the $O(2)$ group, were factored, modulo a
normalisation,  as a
product of the self and anti self dual solutions,
\be
\label{ss1}
{\cal L}=\Big (F^\alpha +\tilde F^\alpha\Big) 
\Big (F^\alpha -\tilde F^\alpha\Big)\,\,\,;\,\,\,D=2k+2
\ee
where the internal index has been explicitly written and the
generalised Hodge operation was defined in (\ref{ssi2}). The key ingredient
in this construction was 
to provide a general definition of self duality that was applicable
for either odd or even $N$ forms. Self duality was now defined to
include the internal space and was implemented either by the
$\sigma_1$ or the $\epsilon$, depending on the dimensionality. 
This naturally led to the universal structure (\ref{ss1}).

Some other aspects of the analysis deserve attention. Specifically, the  
novel  duality symmetric actions obtained in two dimensions
revealed the interpolating role between duality and chirality.
Furthermore, certain points concerning the
interpretation of chirality symmetric action as the
analogue of the duality symmetric electromagnetic action in four
dimensions were clarified.
We also recall that the
soldering of actions to obtain a master action was an intrinsically
quantum phenomenon that
could be expressed in terms of an identity relating two ``gauge variant"
actions to a ``gauge invariant" form. The gauge invariance is with regard
to the set of transformations induced for effecting the soldering and has
nothing to  do with the conventional gauge transformations. In fact the
important thing is that the distinct actions must possess the self and
anti self dual aspects of some symmetry which are being soldered. The
identities obtained in this way are effectively a
generalisation of the usual Polyakov Weigman identity. 
We conclude by stressing
the practical nature of our approach to duality which can be extended to
other theories.  This will be the object of the next section.

\bigskip

\section{Bosonisation and Soldering of Dual Symmetries in
Two and Three Dimensions}

Bosonisation is a powerful technique that  maps a
fermionic theory into its bosonic counterpart. 
It was initially developed and fully
explored in the context of two dimensions\cite{AAR}. More recently,
it has been
extended to higher dimensions\cite{M,C,RB,RB1}.
The importance of bosonisation lies in
the fact that it includes quantum effects already at the classical level. 
Consequently, different aspects and manifestations of quantum phenomena
may be investigated directly, that would otherwise be highly nontrivial
in the fermionic language. Examples of such applications are the 
computation of the current algebra\cite{RB} and the study of screening or
confinement in gauge theories\cite{AB}.

This section is devoted to analyse certain features and applications of
bosonisation which, as far as we are aware, are unexplored even in two
dimensions. The question we pose is the following: given two independent
fermionic models which can be bosonised separately, under what 
circumstances is it possible to represent them by one single effective theory? 
The answer lies in the symmetries of the problem. Two  distinct models
displaying dual aspects of some symmetry can be combined by the
simultaneous implementation of bosonisation and soldering to yield a
completely new theory. This is irrespective of dimensional considerations.
The technique of soldering essentially comprises in lifting the gauging
of a global symmetry to its local version and exploits certain concepts
introduced in a different context by Stone\cite{S} and one of us\cite{W}.
The analysis is intrinsically
quantal without having any classical analogue. This is easily explained
by the observation that a simple addition of two independent classical
lagrangeans is a trivial operation without leading to anything meaningful
or significant.

The basic notions and ideas are first introduced in the context of two
dimensions where bosonisation is known to yield exact results. The 
starting point is to take two distinct chiral lagrangeans with opposite
chirality. Using their bosonised expressions, the soldering mechanism
fuses, in a precise way, the left and right chiralities. This leads to
a general lagrangean in which the chiral symmetry no longer exists, but
it contains  two extra parameters manifesting the bosonisation ambiguities.
It is shown that different  parametrisations lead to different
models. In particular, the gauge invariant Schwinger model and Thirring
model are reproduced. As a byproduct, the importance of Bose symmetry
is realised and some interesting consequences regarding the arbitrary
parametrisation in the chiral Schwinger model are charted.

Whereas the two dimensional analysis lays the foundations, the subsequent
three dimensional discussion illuminates the full power and utility of
the present approach. While the bosonisation in these dimensions is not
exact, nevertheless, for massive fermionic models in the large mass or,
equivalently, the long wavelength limit, well defined local expressions
are known to exist\cite{C,RB}. Interestingly, these expressions exhibit a self
or an anti self dual symmetry that is dictated by the signature of the fermion
mass. Clearly, therefore, this symmetry simulates the dual aspects of
the left and right chiral symmetry in the two dimensional example,
thereby providing a novel testing ground for our ideas. Indeed, two distinct
massive Thirring models with opposite mass signatures,
are soldered to yield a massive Maxwell theory. This result
is vindicated by a direct comparison of the current correlation functions
obtained before and after the soldering process. As another instructive
application, the fusion of two models describing quantum electrodynamics
in three dimensions is considered. Results similar to the corresponding
analysis for the massive Thirring models are obtained.

We conclude by discussing future prospects and possibilities of extending
this analysis in different directions.

\bigskip

\subsection{The two dimensional example}

\bigskip

In this section we develop the ideas in the context of  two dimensions.
Consider, in particular, the following lagrangeans with opposite chiralities, 
\begin{eqnarray}
{\cal L}_+&=&\bar\psi(i \dslash + e \aslash P_+)\psi\nonumber\\
{\cal L}_-&=&\bar\psi(i \dslash + e \aslash P_-)\psi
\label{10}
\end{eqnarray}
where $P_\pm$ are the projection operators,
\begin{equation}
P_\pm=\frac{1 \pm \gamma_5}{2}
\label{20}
\end{equation}
It is well known that the computation of the fermion determinant, which
effectively yields the bosonised expressions, is plagued by regularisation
ambiguities since chiral gauge symmetry cannot be preserved\cite{RJ}. Indeed an
explicit one loop calculation following Schwinger's point splitting method
\cite{RB2} yields,
\begin{eqnarray}
\label{30}
W_+[\varphi] &=&-i \log \det (i\dslash+e\aslash_+)= 
{1\over{4\pi}}\int d^2x\,\left(\partial_+
\varphi\partial_-\varphi +2 \, e\,A_+\partial_-\varphi + a\, 
e^2\, A_+ A_-\right)\nonumber\\
W_-[\rho]&=&-i \log \det (i\dslash+e\aslash_-)= 
{1\over{4\pi}}\int d^2x\,\left(\partial_+\rho\partial_-
\rho +2 \,e\, A_-\partial_+\rho
+ b\, e^2\, A_+ A_-\right)
\end{eqnarray}
where the light cone metric has been invoked for convenience,
\begin{equation}
\label{35}
A_\pm = {1\over\sqrt 2}(A_0\pm A_1)=A^\mp \;\;\; ;\;\;\; \partial_\pm=
{1\over\sqrt 2}(\partial_0\pm \partial_1)=\partial^\mp
\end{equation}
Note that the regularisation or bosonisation ambiguity is manifested through
the arbitrary parameters $a$ and $b$. The latter ambiguity is particularly
transparent since by using the normal bosonisation dictionary
$\bar\psi i\dslash\psi \rightarrow \partial_+\varphi\partial_-\varphi$
and $\bar\psi\gamma_\mu\psi\rightarrow{1\over\sqrt \pi}\epsilon_{\mu\nu}
\partial^\nu\varphi$ (which holds only for a gauge invariant theory),
the above expressions with $a=b=0$ are easily reproduced from (\ref{10}).

It is crucial to observe that different scalar fields $\phi$ and $\rho$
have been used in the bosonised forms to emphasize the fact that the
fermionic fields occurring in the chiral components are uncorrelated.
It is the soldering process which will abstract a meaningful combination
of these components\cite{ABW}. This process essentially consists in lifting the
gauging of a global symmetry to its local version. Consider, therefore,
the gauging of the following global symmetry,
\begin{eqnarray}
\label{40}
\delta \varphi &=& \delta\rho=\alpha\nonumber\\
\delta A_{\pm}&=& 0
\end{eqnarray}

\noindent The variations in the effective actions  (\ref{30}) are found to be,

\begin{eqnarray}
\label{50}
\delta W_+[\varphi] &=& \int d^2x\, \partial_-\alpha \;J_+
(\varphi)\nonumber\\
\delta W_-[\rho]&=& \int d^2x\, \partial_+\alpha \;J_-(\rho)
\end{eqnarray}

\noindent  where the currents are defined as,

\begin{equation}
\label{60}
J_\pm(\eta)={1\over{2\pi}}(\partial_\pm\eta +\, e\,A_\pm)\;\;\; ; \;\;\eta=
\varphi , \rho
\end{equation}

\noindent  The important step now is to introduce the 
soldering field $B_\pm$ coupled with the currents so that,

\begin{equation}
\label{70}
W_\pm^{(1)}[\eta] = W_\pm[\eta] -\int d^2x\, B_\mp\, J_\pm(\eta)
\end{equation}

\noindent Then it is possible to define a modified action,

\begin{equation}
\label{80}
W[\varphi,\rho]= W_+^{(1)}[\varphi] + W_-^{(1)}[\rho]
 + {1\over{2\pi}} \int d^2x \, B_+ \,B_-
\end{equation}

\noindent which is invariant under an extended set of transformations that 
includes (\ref{40}) together with,

\begin{equation}
\delta B_{\pm}= \partial_{\pm}\alpha
\label{90}
\end{equation}

\noindent  Observe that the soldering field transforms exactly as a potential.
It has served its purpose of fusing the two chiral components. Since it 
is an auxiliary field, it can be eliminated from the invariant action 
(\ref{80}) by using the equations of motion. This will naturally solder the
otherwise independent chiral components and justifies its name as a soldering
field. The relevant solution is found to be,

\begin{equation}
\label{100}
B_\pm= 2\pi J_\pm
\end{equation}

\noindent Inserting this solution in (\ref{80}), we obtain,

\begin{equation}
\label{110}
W[\Phi]={1\over {4\pi}}\int d^2x\:\Big{\{}\Big{(}\partial_+
\Phi\partial_-\Phi + 2\,e\, A_+\partial_-\Phi - 2\,e\, A_-
\partial_+\Phi\Big{)} +(a+b-2)\,e^2\,A_+\,A_-\Big{\}}
\end{equation}

\noindent where,

\begin{equation}
\label{120}
\Phi=\varphi - \rho
\end{equation}

\noindent As announced, the action is no longer expreessed in terms of the
different scalars $\varphi$ and $\rho$, but only on their specific combination. 
This combination is gauge invariant. 

Let us digress on the significance of the findings. At the classical fermionic
version, the chiral lagrangeans are completely independent. Bosonising them
includes quantum effects, but still there is no correlation. The soldering
mechanism exploits the symmetries of the independent actions to
precisely combine them to yield a single action.
Note that the soldering works with the bosonised expressions. Thus the soldered
action obtained in this fashion corresponds to the quantum theory.    

We now show that different choices for the parameters $a$ and $b$
 lead to well known models.  To do this consider the variation of
(\ref{110}) under the conventional gauge transformations,
$\delta\varphi=\delta\rho=\alpha$ and $\delta A_\pm = \partial_\pm\alpha$.
It is easy to see that the expression
in parenthesis is gauge invariant. Consequently
a gauge invariant
structure for $W$ is obtained provided,
\begin{equation}
\label{130}
a+b-2=0
\end{equation}

The effect of soldering, therefore, has been to induce a lift of the initial
global symmetry (\ref{40}) to its local form. By functionally integrating
out the $\Phi$ field from (\ref{110}), we obtain,
\begin{equation}
\label{140}
W[A_+,A_-]=  -{ e^2\over 4\pi} \int d^2x\: \{A_+ 
{\partial_-\over \partial_+}A_+ 
+ A_- {\partial_+\over \partial_-}A_- - 2 A_+ A_-\}
\end{equation} 
which is the familiar structure for the gauge invariant action expressed in
terms of the potentials. The opposite chiralities of the 
two independent fermionic theories have been soldered to yield a gauge 
invariant action.

Some interesting observations are possible concerning the regularisation
ambiguity manifested by the parameters $a$ and $b$. Since a single equation
(\ref{130}) cannot fix both the parameters, it might appear that there 
is a whole one parameter class of
solutions for the chiral actions that combine to 
yield the vector gauge invariant
action.  Indeed, without any further input, this is the only conclusion.
However, Bose symmetry imposes a crucial restriction. This symmetry plays
an essential part that complements gauge invariance. Recall, for instance,
the calculation of the triangle graph leading to the Adler-Bell-Jackiw
anomaly. The familiar form of the anomaly cannot be obtained by simply
demanding gauge invariance; Bose symmetry at the vertices of the triangle
must also be imposed\cite{LR,SA}. Similarly, Bose symmetry\cite{RB3}
is essential in reproducing
the structure of the one-cocycle that is mandatory in the analysis on
smooth bosonisation\cite{DNS}; gauge invariance alone fails. In the present case,
this symmetry corresponds to the left-right (or + -) symmetry in (\ref{30}),
thereby requiring $a=b$. Together with the condition (\ref{130}) this implies
$a=b=1$. This parametrisation has important consequences if a Maxwell term
was included from the beginning to impart dynamics. Then the soldering takes
place among two chiral Schwinger models\cite{JR} having opposite chiralities to
reproduce the usual Schwinger model\cite{J}. It
is known that the chiral models satisfy unitarity provided $a, b \geq 1$ and
the spectrum consists of a vector boson with mass,
\be
m^2 = \frac{e^2 a^2}{a-1}
\label{150}
\ee
and a massless chiral boson.  The values of the parameters obtained here
just saturate the bound. 
In other words, the chiral Schwinger model may have any 
$a\geq 1$, but if two such models with opposite chiralities are soldered to 
yield the vector Schwinger model, then the minimal bound is the unique
choice.
Moreover, for the minimal parametrisation,
the mass of the vector boson 
becomes infinite so that it goes out of the spectrum. Thus the
soldering mechanism shows how the massless modes in the chiral Schwinger
models are fused to generate the massive mode of the Schwinger model.

Naively it may appear that the soldering of the left and right chiralities
to obtain a gauge invariant result is a
simple issue since adding the classical lagrangeans
$\bar\psi\Dslash_+\psi$ and $\bar\psi\Dslash_-\psi$, with identical
fermion species, just yields the
usual vector lagrangean $\bar\psi\Dslash\psi$. The quantum considerations 
are, however, much involved. The chiral determinants, as they occur,
cannot be even defined
since the kernels map from one chirality to the other so that there is no
well defined eigenvalue problem\cite{AW,RB3}. This is circumvented by
working with
$\bar\psi(i\dslash + e\aslash_{\pm})\psi$, that satisfy an eigenvalue
equation, from which their determinants may be computed. But now a simple
addition of the classical lagrangeans does not reproduce the expected
gauge invariant form. At this juncture, the soldering process becomes
important. It systematically combined the quantised (bosonised)
expressions for the opposite chiral components. Note that {\it different}
fermionic species were considered so that this soldering does not have
any classical analogue, and is strictly a quantum phenomenon. This will 
become more transparent when the three dimensional case is discussed.

It is interesting to show that a different choice for the parameters $a$
and $b$ in (\ref{110}) leads to the Thirring model. Indeed it is precisely
when the mass term exists ($i.e., \,\, a+b-2\neq 0$), that (\ref{110})
represents
the Thirring  model. Consequently, this parametrisation complements that used
previously to obtain the vector gauge invariant structure. It is now easy to
see that the term in parentheses in (\ref{110}) corresponds to $\bar\psi
(i\dslash +e\aslash) \psi$ so that integrating out the auxiliary $A_\mu$ field
yields,
\be
{\cal L}=\bar\psi i\dslash\psi - \frac{g}{2}(\bar\psi\gamma_\mu\psi)^2\,\,\,\,
;\, g=\frac{4\pi}{a+b-2}
\label{A1}
\ee
which is just the lagrangean for the usual Thirring model. It is known
\cite{SC}that 
this model is meaningful provided the coupling parameter satisfies the 
condition $g>-\pi$, so that,
\be
\label{A2}
\mid a+b \mid >2
\ee
This condition is the analogue of (\ref{130}) found earlier. As usual, there
is a one parameter arbitrariness. Imposing Bose symmetry implies that both
$a$ and $b$ are equal and lie in the range
\be
\label{A3}
1<\mid a\mid =\mid b\mid
\ee
This may be compared with the previous case where $a=b=1$ was necessary for
getting the gauge invariant structure.  Interestingly, the positive range
for the parameters in (\ref{A3}) just commences from this value.

Having developed and exploited the concepts of soldering in two dimensions,
it is natural to investigate their consequences in three dimensions. The
discerning reader may have noticed that it is essential to have dual
aspects of a symmetry that can be soldered to yield new information. In the
two dimensional case, this was the left and right chirality. Interestingly,
in three dimensions also, we have a similar phenomenon.

\bigskip
\bigskip

\subsection{The three dimensional example}

\bigskip

This section is devoted to an analysis of the soldering process in the massive
Thirring model in three dimensions.  We shall show that two apparently
independent massive Thirring models in the long wavelength limit combine,
at the quantum level, into a massive Maxwell theory. 
This is further vindicated by a direct comparison of the current correlation
functions following from the bosonization identities. 
These findings are also extended to include three dimensional
quantum electrodynamics. 
The new results and interpretations illuminate a close
parallel with the two dimensional discussion.

\bigskip

\subsection{The massive Thirring model}

\bigskip

In order to effect the soldering, the first step is to consider the
bosonisation of the massive Thirring
model in three dimensions\cite{C,RB}. This is therefore reviewed briefly. The
relevant current correlator generating functional,
in the Minkowski metric, is given by,
\be
\label{160}
Z[\kappa]=\int D\psi D\bar\psi \exp\Bigg(i\int d^3x\Bigg
[\bar\psi(i \dslash + m )\psi -\frac{\lambda^2}{2}
j_\mu j^\mu + \lambda j_\mu \kappa^\mu\Bigg]\Bigg)
\ee
where $j_\mu=\bar\psi\gamma_\mu\psi$ is the fermionic current. As usual,
the four fermion interaction can be eliminated by introducing an auxiliary
field,
\be
\label{170}
Z[\kappa]=\int D\psi D\bar\psi Df_\mu\exp\Bigg(i\int d^3x\Bigg
[\bar\psi\left(i \dslash + m +\lambda (\fslash +\kslash)\right)
\psi +\frac{1}{2} f_\mu f^\mu\Bigg]\Bigg)
\ee
Contrary to the two dimensional models, the fermion integration cannot be
done exactly. Under certain limiting conditions, however, this integation
is possible leading to closed expressions. A particularly effective choice
is the large mass limit in which case the fermion determinant yields a local
form. Incidentally, any other value of the mass leads to a nonlocal structure
\cite{RB1}.
The large mass limit is therefore very special. The leading term in this
limit was calculated by various means \cite{DJT} 
and shown to yield the Chern-Simons
three form. Thus the generating functional for the massive Thirring model in
the large mass limit is given by,
\be 
\label{180}
Z[\kappa]=\int Df_\mu 
\exp\Bigg( i\int d^3x\:\Bigg({\lambda^2\over{8\pi}}{m\over{\mid m\mid}}
\epsilon_{\mu\nu\lambda}f^\mu\partial^\nu f^\lambda +
\frac{1}{2} f_\mu f^\mu +\frac{\lambda^2}{4\pi}\frac{m}{\mid m\mid}
\epsilon_{\mu\nu\sigma}\kappa^\mu\partial^\nu f^\sigma\Bigg)\Bigg)
\ee
where the signature of the topological terms is dictated by the corresponding
signature of the fermionic mass term. 
In obtaining the above result a local counter term has been ignored. 
Such terms manifest the ambiguity in defining the time ordered product
to compute the correlation functions\cite{BRR}.
The lagrangean in the above partition
function defines a self dual model introduced earlier \cite{TPN}. The massive
Thirring model, in the relevant limit, therefore bosonises to a self dual
model. It is useful to clarify the meaning of this self duality. The 
equation of motion in the absence of sources is given by,
\be
\label{190}
f_\mu =-{\lambda^2\over{4\pi}}{m\over{\mid m\mid}}
\epsilon_{\mu\nu\lambda}\partial^\nu f^\lambda  
\ee
from which the following relations may be easily verified,
\br
\label{200}
\partial_\mu f^\mu &=& 0\nonumber\\
\left(\Box + M^2\right)f_\mu &=& 0 \,\,\,\,\,\, ;\,\, 
M=\frac{4\pi}{\lambda^2}
\er
A field dual to $f_\mu$ is defined as,
\be
\label{210}
\tilde f_\mu = {1\over M} \epsilon_{\mu\nu\lambda}\partial^\nu f^\lambda
\ee
where the mass parameter $M$ is inserted for dimensional reasons. Repeating
the dual operation, we find,
\be
\label{220}
\tilde{\left(\tilde{f_\mu}\right)}= 
{1\over M} \epsilon_{\mu\nu\lambda}\partial^\nu\tilde{f^\lambda}=f_\mu
\ee
obtained by exploiting (\ref{200}), thereby validating the definition of
the dual field.  Combining these results with (\ref{190}),
we conclude that,
\be
\label{230}
f_\mu=- \frac{m}{\mid m \mid} \tilde f_\mu
\ee
Hence, depending on the sign of the fermion mass term, the bosonic theory
corresponds to a self-dual or an anti self-dual model.  Likewise, the Thirring
current bosonises to the topological current

\be
\label{235}
j_\mu = \frac{\lambda}{4\pi}\frac{m}{\mid m\mid}\epsilon_{\mu\nu\rho}
\partial^\nu f^\rho
\ee

The close connection with the two dimensional analysis is now evident.
There the starting point was to consider two distinct fermionic theories with 
opposite chiralities. In the present instance, the analogous thing is to
take two independent Thirring models with identical coupling strengths but
opposite mass signatures,
\br
\label{240}
{\cal L_+}&=&\bar\psi\left(i\dslash + m\right)\psi -\frac{\lambda^2}{2}\left(\bar\psi\gamma_\mu\psi\right)^2\nonumber\\
{\cal L_-} &=& \bar \xi\left(i\dslash - m'\right)\xi - \frac{\lambda^2}{2} \left(\bar\xi\gamma_\mu\xi\right)^2
\er
Note that the only the relative sign between the mass parameters is important,
but their magnitudes are different. From now on it is also assumed that
both $m$ and $m'$ are positive.
Then the bosonised lagrangeans are, respectively,
\br
\label{250}
{\cal L_+}&=&\frac{1}{2M} 
\epsilon_{\mu\nu\lambda}f^\mu\partial^\nu f^\lambda +
{1\over 2} f_\mu f^\mu\nonumber\\
{\cal L_-} &=&- \frac{1}{2M}
\epsilon_{\mu\nu\lambda}g^\mu\partial^\nu g^\lambda +
{1\over 2} g_\mu g^\mu
\er
where $f_\mu$ and $g_\mu$ are the distinct bosonic vector fields.  The current
bosonization formulae in the two cases are given by

\br
\label{255}
j_\mu^+ &=& \bar\psi\gamma_\mu\psi=\frac{\lambda}{4\pi} 
\epsilon_{\mu\nu\rho}\partial^\nu f^\rho\nonumber\\
j_\mu^- &=&\bar\xi\gamma_\mu\xi= - \frac{\lambda}{4\pi} 
\epsilon_{\mu\nu\rho}\partial^\nu g^\rho
\er

The stage is now set for soldering. Taking a cue from the two dimensional
analysis, let us consider the gauging of the following symmetry,
\be
\label{260}
\delta f_\mu = \delta g_\mu = 
\epsilon_{\mu\rho\sigma}\partial^\rho \alpha^\sigma
\ee
Under such transformations, the bosonised lagrangeans change as,
\be
\label{270}
\delta{\cal L_\pm} = J_\pm^{\rho\sigma}(h_\mu)
 \partial_\rho\alpha_\sigma \,\,\,\,\, ;\,\, h_\mu=f_\mu,\,\,g_\mu
\ee
where the antisymmetric currents are defined by,
\be
\label{280}
J_\pm^{\rho\sigma}(h_\mu)= \epsilon^{\mu\rho\sigma}h_\mu \pm {1\over M}
\epsilon^{\gamma\rho\sigma}\epsilon_{\mu\nu\gamma}\partial^\mu 
h^\nu
\ee
It is worthwhile to mention that any other variation of the fields
(like $\delta{f_\mu}=\alpha_\mu$)is inappropriate because changes in
the two terms of the lagrangeans cannot be combined to give a single
structure like (\ref{280}). We now introduce the soldering field coupled
with the antisymmetric currents. In
the two dimensional case this was a vector. Its natural extension now
is the antisymmetric second rank Kalb-Ramond tensor field $B_{\rho\sigma}$,
transforming in the usual way,
\be
\label{290}
\delta B_{\rho\sigma}=\partial_\rho\alpha_\sigma -
\partial_\sigma\alpha_\rho
\ee
Then it is easy to see that the modified lagrangeans,
\be
\label{300}
{\cal L}_\pm^{(1)}={\cal L}_\pm - {1\over 2} J_\pm^{\rho\sigma}(h_\mu)
B_{\rho\sigma}
\ee
transform as,
\be
\label{310}
\delta{\cal L}_\pm^{(1)}=- {1\over 2} \delta J_\pm^{\rho\sigma}
B_{\rho\sigma}
\ee
The final modification consists in adding a term to ensure gauge invariance
of the soldered lagrangean. This is achieved by,
\be
\label{320}
{\cal L}_\pm^{(2)}={\cal L}_\pm^{(1)} + {1\over 4} 
B^{\rho\sigma}B_{\rho\sigma}
\ee
A straightforward algebra shows that the following combination,
\br
\label{330}
{\cal L}_S &=& {\cal L}_+^{(2)}+{\cal L}_-^{(2)}\nonumber\\
&=&{\cal L}_+ + {\cal L}_- -{1\over 2}B^{\rho\sigma}
\left(J^+_{\rho\sigma}(f) + J^-_{\rho\sigma}(g)\right)
+{1\over 2} B^{\rho\sigma}B_{\rho\sigma}
\er
is invariant under the gauge transformations (\ref{260}) and (\ref{290}). 
The gauging of the symmetry
is therefore complete. To return to a description in terms of the original
variables, the auxiliary soldering field is eliminated from (\ref{330}) by 
using the equation of motion,
\be
\label{340}
B_{\rho\sigma}= {1\over 2} \left(J_{\rho\sigma}^+(f)+
J_{\rho\sigma}^-(g)\right)
\ee
Inserting this solution in (\ref{330}), the final soldered
lagrangean is expressed
solely in terms of the currents involving the original fields,
\be
\label{350}
{\cal L}_S ={\cal L}_+ + {\cal L}_- -
{1\over 8}\left(J_{\rho\sigma}^+(f)+
J_{\rho\sigma}^-(g)\right)\left(J^{\rho\sigma}_+(f)+
J^{\rho\sigma}_-(g)\right)
\ee
It is now crucial to note that, by using the explicit structures for the
currents, the above lagrangean is no longer a function of $f_\mu$ and $g_\mu$
separately, but only on the combination,
\be
\label{360}
A_\mu = {1\over{\sqrt{2} M}}\left(g_\mu - f_\mu\right)
\ee
with,
\be
\label{370}
{\cal L}_S = - \frac{1}{4} F_{\mu\nu}F^{\mu\nu} + 
{M^2\over 2}A_\mu A^\mu
\ee
where,
\be
\label{380}
F_{\mu\nu}= \partial_\mu A_\nu -\partial_\nu A_\mu
\ee
is the usual field tensor expressed in terms of the basic entity $A_\mu$.
Our goal has been achieved. The soldering mechanism has precisely fused
the self and anti self dual symmetries to yield a massive Maxwell theory
which, naturally, lacks this symmetry.

It is now instructive to understand this result by comparing the current
correlation functions.  The Thirring currents in the two models bosonise
to the topological currents (\ref{255}) in the dual formulation.  From a
knowledge of the field correlators in the latter case, it is therefore
possible to obtain the Thirring current correlators.  The field
correlators are obtained from the inverse of the kernels occurring in
(\ref{250}),

\br
\label{381}
\langle f_\mu(+k)\: f_\nu(-k)\rangle &=&
\frac{M^2}{M^2 - k^2}\left(i g_{\mu\nu} +
{1\over M}\epsilon_{\mu\rho\nu} k^\rho -
\frac{i}{M^2} k_\mu\:k_\nu\right)\nonumber\\
\langle g_\mu(+k)\: g_\nu(-k)\rangle &=&
\frac{M^2}{M^2 - k^2}\left(i g_{\mu\nu} -
{1\over M}\epsilon_{\mu\rho\nu} k^\rho -
\frac{i}{M^2} k_\mu\:k_\nu\right)
\er

\noindent where the expressions are given in the momentum space.
Using these in (\ref{255}), the current
correlators are obtained, 

\br
\label{382}
\langle j_\mu^+(+k) j_\nu^+(-k)\rangle &=&
\frac{M}{4\pi (M^2 - k^2)}
\left(i k^2 g_{\mu\nu}- ik_\mu\: k_\nu
+{1\over M}\epsilon_{\mu\nu\rho}
k^\rho\:k^2\right)\nonumber\\
\langle j_\mu^-(+k) j_\nu^-(-k)\rangle &=&
\frac{M}{4\pi (M^2 - k^2)}
\left(i k^2 g_{\mu\nu}- ik_\mu\: k_\nu
-{1\over M}\epsilon_{\mu\nu\rho}
k^\rho\:k^2\right)
\er
It is now feasible to construct a total current,

\be
\label{383}
j_\mu=j_\mu^+ + j_\mu^- = \frac{\lambda}{4\pi}
\epsilon_{\mu\nu\rho}\partial^\nu 
\left(f^\rho - g^\rho\right)
\ee
Then the correlation function for this current, in the original self dual
formulation, follows
from (\ref{382}) and noting that $\langle j_\mu^+\:
j_\nu^-\rangle =0$, which is a consequence of the
independence of $f_\mu$ and $g_\nu$;

\be
\label{384}
\langle j_\mu(+k)\: j_\nu(-k)\rangle = \langle j_\mu^+\: j_\nu^+\rangle +
\langle j_\mu^-\: j_\nu^-\rangle =
\frac{iM}{2\pi(M^2 -k^2)}
\left(k^2\: g_{\mu\nu} - k_\mu\: k_\nu\right)
\ee
The above equation is easily reproduced from the 
effective theory.  Using (\ref{360}), it is observed that the
bosonization of the composite current (\ref{383}) is
defined in terms of the massive vector field $A_\mu$,

\be
\label{385}
j_\mu=\bar\psi\gamma_\mu\psi +\bar\xi\gamma_\mu\xi =
-\sqrt{{M\over 2\pi}}\epsilon_{\mu\nu\rho}
\partial^\nu A^\rho
\ee
The current correlator is now obtained from the field
correlator $\langle A_\mu\: A_\nu\rangle$ given by the
inverse of the kernel appearing in (\ref{370}),

\be
\label{386}
\langle A_\mu(+k)\: A_\nu(-k)\rangle =
\frac{i}{M^2 - k^2}\left(g_{\mu\nu} - 
\frac{k_\mu\: k_\nu}{M^2}\right)
\ee
>From (\ref{385}) and (\ref{386}) the two point function
(\ref{384}) is reproduced, including the normalization.

We conclude, therefore, that two massive Thirring models with opposite 
mass signatures, in the long wavelength limit,
combine by the process of bosonisation and soldering, to a massive
Maxwell theory. The bosonization of the composite current, obtained
by adding the separate contributions from the two models, is given in
terms of a topological current(\ref{385}) of the massive vector theory.
These are completely new results which cannot be obtained by a
straightforward application of conventional bosonisation techniques.
The massive modes in the original Thirring models are 
manifested in the two modes of (\ref{370}) so that there is a proper
matching in the degrees of freedom.
Once again it is reminded that the 
fermion fields for the models are different so that the analysis has no
classical analogue. Indeed if one considered the same fermion species,
then a simple addition of the classical lagrangeans would lead to a 
Thirring model with a mass given by $m-m'$. 
In particular, this difference can be zero.
The bosonised version of such a massless model is
known \cite{M, RB1} to yield a highly nonlocal theory which has no connection
with (\ref{370}). Classically, therefore, there is no possibility of even
understanding, much less, reproducing the effective quantum result. 
In this sense the application in three dimensions
is more dramatic than the corresponding case of two dimensions.
        
\bigskip

\subsection{Quantum electrodynamics}

\bigskip

An interesting theory in which the preceding ideas may be
implemented is quantum electrodynmics, whose current
correlator generating
functional in an arbitrary covariant gauge is given by,
\be
\label{390}
Z[\kappa]=\int D\bar\psi\: D\psi\: DA_\mu\, \exp
\left\{i\int d^3x\:\left(\bar\psi\left(i\dslash + m +e\aslash
\right)\psi -{1\over 4} F_{\mu\nu}F^{\mu\nu} 
+{\eta\over 2}(\partial_\mu A^\mu)^2 + e j_\mu\kappa^\mu
\right)\right\}
\ee
where $\eta$ is the gauge fixing parameter and
$j_\mu = \bar\psi\gamma_\mu\psi$ is the current. 
As before, a one loop computation of the fermionic
determinant in the large mass limit yields,
\br
\label{400}
Z[\kappa]&=&\int  DA_\mu\, 
\exp { \lbrace} i\int d^3x\:\lbrack\frac{e^2}{8\pi}
\frac{m}{\mid m\mid}
\epsilon_{\mu\nu\lambda}A^\mu\partial^\nu A^\lambda
-{1\over 4} F_{\mu\nu}F^{\mu\nu}\nonumber\\
&+& \frac{e^2}{4\pi}
\frac{m}{\mid m \mid} \epsilon_{\mu\nu\rho}
\kappa^\mu\partial^\nu\: A^\rho +{\eta\over 2}
(\partial_\mu A^\mu)^2\rbrack{\rbrace}
\er
In the absence of sources, this just corresponds to the
topolologically massive Maxwell-Chern-Simons theory, with
the signature of the topological term determined from that of
the fermion mass term. The equation of motion,

\be
\label{405}
\partial^\nu\, F_{\nu\mu} +\frac{e^2}{4\pi}
\frac{m}{\mid m \mid} \epsilon_{\mu\nu\lambda}
\partial^\nu A^\lambda = 0
\ee
expressed in terms of the dual tensor,

\be
\label{410}
F_\mu = \epsilon_{\mu\nu\lambda}
\partial^\nu A^\lambda
\ee
reveals the self (or anti self) dual property,
\be
\label{420}
F_\mu =\frac{4\pi}{e^2}\frac{m}{\mid m\mid}
\epsilon_{\mu\nu\lambda}\partial^\nu F^\lambda
\ee
which is the analogue of (\ref{190}).  In this fashion the Maxwell-Chern-Simons theory
manifests the well known \cite{DJ, BRR, BR} mapping with the self
dual models considered in the previous subsection.
The difference is that the self duality in the former,
in contrast to the latter, is contained in the dual
field (\ref{410}) rather than in the basic field defining
the theory. This requires some modifications in the
ensuing analysis.  Furthermore, the bosonization of the
fermionic current is now given by the topological current
in the Maxwell-Chern-Simons theory,

\be
\label{425}
j_\mu = \frac{e}{4\pi} \frac{m}{\mid m \mid}
\epsilon_{\mu\nu\lambda}\partial^\nu A^\lambda
\ee

Consider, therefore, two independent models describing quantum 
electrodynamics with opposite signatures in the mass terms,

\br
\label{426}
{\cal L}_+ &=& \bar\psi\left( i\dslash +m
+e\aslash\right)\psi -{1\over 4} 
F_{\mu\nu}(A) F^{\mu\nu}(A)\nonumber\\
{\cal L}_- &=& \bar\xi\left( i\dslash -m'
+e\bslash\right)\xi -{1\over 4} 
F_{\mu\nu}(B) F^{\mu\nu}(B)
\er

whose bosonised versions in an appropriate limit are given by,
\br
\label{430}
{\cal L}_+ &=& -\frac{1}{4} F_{\mu\nu}(A)+\frac{M}{2}
\epsilon_{\mu\nu\lambda}A^\mu\partial^\nu A^\lambda 
\,\,\,\,\, ; \,\, M=\frac{e^2}{4\pi}\nonumber\\
{\cal L}_- &=& -\frac{1}{4} F_{\mu\nu}(B)-\frac{M}{2}
\epsilon_{\mu\nu\lambda}B^\mu\partial^\nu B^\lambda
\er
where $A_\mu$ and $B_\mu$ are the corresponding potentials. 
Likewise, the corresponding expressions for the bosonized
currents are found from the general structure (\ref{425}),

\br
\label{435}
j_\mu^+ &=&\bar\psi\gamma_\mu\psi= \frac{M}{e} \epsilon_{\mu\nu\lambda}
\partial^\nu A^\lambda\nonumber\\
j_\mu^- &=&\bar\xi\gamma_\mu\xi= -\frac{M}{e} \epsilon_{\mu\nu\lambda}
\partial^\nu B^\lambda
\er
To proceed with
the soldering of the above models, take  the symmetry transformation,
\be
\label{440}
\delta A_\mu=\alpha_\mu
\ee
Such a transformation is spelled out by recalling (\ref{260}) and the 
observation that now (\ref{410}) simulates the $f_\mu$ field in the previous
case. Under this variation, the lagrangeans (\ref{430}) change as,
\be
\label{450}
\delta{\cal L}_\pm =J_\pm^{\rho\sigma}(P)\partial_\rho\alpha_\sigma
\,\,\,\,\, ; \,\, P=A,B
\ee
where the antisymmetric  currents are defined by,
\be
\label{460}
J_\pm^{\rho\sigma}(P)=\pm m \epsilon^{\rho\sigma\mu}P_\mu -F^{\rho\sigma}(P)
\ee
Proceeding as before, the antisymmetric soldering field $B_{\alpha\beta}$
transforming as (\ref{290}) is introduced by coupling 
with these currents to define the
first iterated lagrangeans analogous to (\ref{300}),
\be
\label{470}
{\cal L}_\pm^{(1)}={\cal L}_\pm - 
{1\over 2} J_\pm^{\rho\sigma}(P) B_{\rho\sigma}
\ee
These lagrangeans are found to transform as,
\be
\label{480}
\delta{\cal L}_\pm^{(1)} ={1\over 4} \delta B^2_{\lambda\sigma}
-{1\over 2}\left(\pm m \epsilon_{\mu\lambda\sigma}\alpha^\mu
B^{\lambda\sigma}\right)
\ee
It is now straightforward to deduce the final lagrangean that will be
gauge invariant. This is given by,
\be
\label{490}
{\cal L}_S = {\cal L}_+^{(2)} + {\cal L}_-^{(2)}
\,\,\,\,\, ; \,\, \delta{\cal L}_S =0
\ee
where the second iterated pieces are,
\be
\label{500}
{\cal L}_\pm^{(2)}= {\cal L}_\pm -{1\over 2} J_\pm^{\rho\sigma}
B_{\rho\sigma} -{1\over 4}B_{\rho\sigma}B^{\rho\sigma}
\ee
The invariance of ${\cal L}_S$ (\ref{490}) is verified by observing that,
\be
\label{510}
\delta{\cal L}_\pm^{(2)}=\mp{1\over 2}m\epsilon_{\mu\lambda\sigma}
\alpha^\mu B^{\lambda\sigma}
\ee
To obtain the effective lagrangean it is necessary to eliminate the auxiliary
$B_{\rho\sigma}$ field by using the equation of motion following from
(\ref{490}),
\be
\label{520}
B_{\sigma\lambda}=-{1\over 2}\left(J^+_{\sigma\lambda}(A) +
J^-_{\sigma\lambda}(B)\right)
\ee
Putting this back in (\ref{490}), we obtain the final soldered lagrangean,
\be
\label{530}
{\cal L}_S =-{1\over 4} F_{\mu\nu}(G)F^{\mu\nu}(G) 
+\frac{M^2}{2}G_\mu G^\mu
\ee
written in terms of a single field,
\be
\label{540}
G_\mu =\frac{1}{\sqrt{2}}\left(A_\mu - B_\mu\right)
\ee
The lagrangean (\ref{530}) governs the dynamics of a massive Maxwell theory.

As before, we now discuss the implications for the current
correlation functions.  These functions in the original models
describing electrodynamics can be obtained from the mapping
(\ref{435}).  The first step is to abstract the basic field
correlators found by inverting the kernels occurring in
(\ref{430}).  The results, in the momentum space, are

\br
\label{545}
\langle A_\mu(+k)\: A_\nu(-k)\rangle &=&
\frac{i}{M^2 -k^2}\left[ g_{\mu\nu} +
\frac{M^2 -k^2(\eta +1)}{\eta k^4}k_\mu \: k_\nu +
\frac{iM}{k^2} \epsilon_{\mu\rho\nu}k^\rho\right]
\nonumber\\
\langle B_\mu(+k)\: B_\nu(-k)\rangle &=&
\frac{i}{M^2 -k^2}\left[ g_{\mu\nu} +
\frac{M^2 -k^2(\eta +1)}{\eta k^4}k_\mu \: k_\nu -
\frac{iM}{k^2} \epsilon_{\mu\rho\nu}k^\rho\right]
\er
The current correlators are easily computed by
substituting (\ref{545}) into (\ref{435}),

\br
\label{546}
\langle j_\mu^+(+k)\:j_\nu^+(-k)\rangle &=&
i\left(\frac{M}{e}\right)^2\frac{1}{M^2 -k^2}
\left[k^2 g_{\mu\nu} -k_\mu\: k_\nu -
i M\epsilon_{\mu\nu\rho}k^\rho\right]
\nonumber\\
\langle j_\mu^-(+k)\:j_\nu^-(-k)\rangle &=&
i\left(\frac{M}{e}\right)^2\frac{1}{M^2 -k^2}
\left[k^2 g_{\mu\nu} -k_\mu\: k_\nu +
i M\epsilon_{\mu\nu\rho}k^\rho\right]
\er
where, expectedly, the gauge dependent ($\eta$) contribution has
dropped out.  Defining a composite current,

\be
\label{547}
j_\mu = j_\mu^+ + j_\mu^- = \frac{M}{e} 
\epsilon_{\mu\nu\lambda}\partial^\nu
\left(A^\lambda - B^\lambda\right)
\ee
it is simple to obtain the relevant correlator
by exploiting the results for $j_\mu^+$ and $j_\mu^-$ from (\ref{546}),

\be
\label{548}
\langle j_\mu(+k)\: j_\nu(-k)\rangle =
2i\left(\frac{M}{e}\right)^2\frac{1}{M^2 -k^2} 
\left(k^2 g_{\mu\nu} - k_\mu\: k_\nu\right)
\ee
In the bosonized version obtained from the soldering,
(\ref{547}) represents the mapping,

\be
\label{549}
j_\mu=\bar\psi\gamma_\mu\psi +\bar\xi\gamma_\mu\xi =
\sqrt 2 {M\over e} \epsilon_{\mu\nu\lambda}
\partial^\nu G^\lambda
\ee
where $G_\mu$ is the massive vector field (\ref{540})
whose dynamics is governed by the lagrangean (\ref{530}). 
In this effective description the result (\ref{548}) is
reproduced from (\ref{549}) by using the correlator
of $G_\mu$ obtained from (\ref{530}), which is exactly identical to
(\ref{386}).

Thus the combined effects of bosonisation and soldering show that two
independent
quantum electrodynamical models with appropriate mass signatures are
equivalently described by the massive Maxwell theory. In the self dual version
the massive modes are the topological excitations in the Maxwell-Chern-Simons
theories. These are combined into the two usual massive modes in the effective
massive vector theory.
A complete correspondence among the composite current correlation
functions in the original models and in their dual bosonised
description was also established.  The comments made
in the concluding part of the last subsection naturally apply also in this
instance.

\bigskip

\subsection{Final Remarks}

\bigskip

The present analysis clearly revealed the possibility of obtaining new
results from quantum effects that conspire to combine two apparently
independent theories into a single effective theory. 
The essential ingredient was that these theories
must possess the dual aspects of the same symmetry. Then, by a systematic
application of bosonisation and soldering, it was feasible to abstract
a meaningful combination of such models, which can never be obtained
by a naive addition of the classical lagrangeans.

The basic notions and ideas were particularly well illustrated in the
two dimensional example where the bosonised expressions for distinct
chiral lagrangeans were soldered to reproduce either the usual gauge
invariant theory or the Thirring model. Indeed, the soldering mechanism
that fused the opposite chiralities, clarified several aspects of the  
ambiguities occurring in bosonising chiral lagrangeans. It was clearly 
shown that unless Bose symmetry is imposed as an additional restriction,
there is a whole one parameter class of bosonised solutions for the chiral
lagrangeans that can be soldered to yield the vector gauge invariant result.
The close connection between Bose symmetry and gauge invariance was thereby
established, leading to a unique parametrisation. Similarly, using a 
different parametrisation, the soldering of the chiral lagrangeans led to
the Thirring model. Once again there was a one parameter ambiguity unless
Bose symmetry was imposed. If that was done, there was a specified range of 
solutions for the chiral lagrangeans that combined to yield a well defined
Thirring model.

The elaboration of our methods was done by considering the massive version
of the Thirring model and quantum electrodynamics in three dimensions.
By the process of bosonisation such models, in the long wavelength limit, 
were cast in a form which 
manifested a self dual symmetry. This was a basic perquisite for effecting
the soldering. It was explicitly shown that two distinct massive Thirring 
models, with opposite mass signatures, combined to a massive Maxwell theory.
The Thirring current correlation functions calculated either in the original
self dual formulation or in the effective massive vector theory yielded 
identical results, showing the consistency of our approach. The application
to quantum electrodynamics followed along similar lines.

It is evident that the present technique of combining models by the two
step process of bosonisation and soldering can be carried through in
higher dimensions provided the models have the relevant symmetry properties.
It is also crucial to note that duality pervades the entire analysis.
In the three dimensional case this was self evident since the models
had a self (and  anti) self dual symmetry. This was hidden in the two
dimensional case where chiral symmetry was more transparent. But it may be
mentioned  that in two dimensions, chiral symmetry is the analogue of 
the duality $\partial_\mu\phi=\pm\epsilon_{\mu\nu}\partial^\nu\phi$. 
Interestingly, the duality
in two dimensions was manifest in the lagrangeans while that in three 
dimensions was contained in the equations of motion. This opens up
the possibility to discuss different aspects of duality, contained either
in the lagrangean or in the equations of motion, in the same framework.
Consequently, the methods developed here can be relevant and useful in
diferent contexts; particularly in the recent discussions on electromagnetic
duality or the study of chiral forms which exactly possess the type of self
dual symmetry considered in this paper. We will report on these and related
issues in a future work.\vspace{1cm}

{\bf Acknowledgements} We would like to thank Manu Paranjape and the members of the Physics Department in the Universite de Montreal for the kind invitation that made this visit possible.  This author is partially supported by CNPq, FUJB and FAPERJ, Brasilian scientific agencies.

\end{document}